\documentclass[12pt]{article}
\usepackage{epsfig,amsmath,amssymb,amsthm,thmtools,latexsym,graphicx,setspace,fullpage}
\usepackage{chicago} 
\usepackage{rotating}
\usepackage{breqn}
\usepackage{authblk}

\allowdisplaybreaks
\usepackage{color} 

\declaretheoremstyle[
notefont=\bfseries, notebraces={}{},
bodyfont=\normalfont,
postheadspace=0.5em,
numbered=no,
]{mystyle}

\begin{document}

\renewcommand{\thefootnote}{\fnsymbol{footnote}}

\title{\vspace{-5mm}Using history matching for prior choice}

\author[1]{Wang Xueou}
\author[1]{David J. Nott\thanks{Corresponding author:  standj@nus.edu.sg}}
\author[2]{Christopher C. Drovandi}
\author[2,3]{Kerrie Mengersen}
\author[4]{Michael Evans}
\affil[1]{Department of Statistics and Applied Probability, National University of Singapore, Singapore 117546}
\affil[2]{School of Mathematical Sciences, Queensland University of Technology, Australia}
\affil[3]{ACEMS, the ARC Centre of Excellence in Mathematical and Statistical Frontiers}
\affil[4]{Department of Statistics, University of Toronto, Toronto, Ontario, M5S 3G3, Canada}

\date{}

\maketitle

\doublespacing

\vspace*{-11mm}\noindent

\begin{abstract}
\noindent  
It can be important in Bayesian analyses of complex models to construct informative prior distributions which reflect knowledge external to the data at hand.  
Nevertheless, how much prior information an analyst can elicit from an expert will be limited due to constraints of time, cost
and other factors.
This paper develops effective numerical methods for exploring reasonable choices of a prior distribution from a parametric
class, when prior information is specified in the form of some limited constraints on prior predictive distributions, and where these prior predictive
distributions are analytically intractable.  
The methods developed may be thought of as a novel application of the ideas of history matching, 
a technique developed in the literature on assessment of computer models.  
We illustrate the approach in the context of logistic regression and sparse signal shrinkage prior distributions for high-dimensional linear models.  
\vspace{2mm}

\end{abstract}

\section{Introduction}

Elicitation of a prior distribution is an important part of Bayesian analysis.  However, often
a detailed representation of an expert's beliefs is difficult to obtain, assuming it is reasonable to suppose that there are true probabilities representing an expert's beliefs at all.  
Even if it were possible to perform comprehensive elicitations in complex multivariate situations, it might not be worth the cost involved in many cases.  
In complex models, how much prior information can be easily elicited from an expert will be limited due to constraints of time, cost and other factors.
For an overview of modern prior elicitation methods including realistic goals of the process, ways of evaluating its success, and the cognitive biases that make it difficult see \shortciteN{garthwaite+ko05}, \shortciteN{ohagan+bdegjor08}, \shortciteN{daneshkhah+o10},
\shortciteN{martin+bfklmm12}, \shortciteN{simpson+rmrs14} and \shortciteN{morris+oc14}, among others.
For a recent discussion of model checking including criticism of the prior see Chapter 5 of \shortciteN{evans15}.

Here we consider the problem of predictive elicitation, where prior information is given by certain limited constraints
on prior predictive distributions which are not analytically tractable. By limited constraints we mean that the given prior information might rule out
some distributions as unsuitable for the prior, but the prior information does not identify a unique suitable prior distribution.   
We will be concerned with
developing effective numerical methods for finding a reasonable value or set of values for a prior hyperparameter 
so that the prior satisfies the constraints. 
It is not our intention
in this manuscript to consider the best ways to elicit the predictive constraints from an expert - these are assumed to be given - and 
the numerical methods discussed here are a tool to be used as part of an iterative process of questioning and feedback in the elicitation context.  
A more comprehensive discussion of elicitation methods is given in the references above.

The method we propose can be thought of as a novel application of the method of history matching (\shortciteNP{craig+gss97}) used in the literature on assessment
of computer models.  A recent application of history matching in the context of a complex infectious diseases model that describes the history matching approach
is \shortciteN{andrianakis+others15}.  We delay further discussion of the relevant literature to Section 3.  
Computer models, sometimes called ``simulators", are complex computer codes that take certain inputs or parameters and produce an output.  The models can either
be stochastic or deterministic.  The goal of history matching is to eliminate regions of the computer model 
parameter space where predictions from the computer model are clearly inconsistent with observed data.  This may result in the conclusion that there 
are no plausible values of the parameters given the level of model discrepancy considered to be reasonable, and the results of a history match can guide model
development and make any subsequent calibration of the model more efficient.  

To apply history matching to the problem of prior choice, we can 
view the prior hyperparameters as the computer model
parameters, and 
use characteristics of the prior predictive densities as the computer model outputs.
From these outputs an implausibility measure of the type used in history matching can be constructed.  
Similar to the computer models context, the approach can give an indication that there are no priors within the class considered satisfying the stated
predictive constraints, as well as exploring the set of possible prior choices when the set of constraints allow for a number of suitable priors.  
The set of appropriate prior choices returned by the method can be used as a basis for making a unique prior choice less arbitrary, as a starting point
for adding further information, or in a sensitivity analysis.  

The method we discuss here, while focusing on computational problems, 
is in the tradition of predictive elicitation methods which elicit information about potentially observable data, 
rather than eliciting information about parameters directly.  Examples of predictive elicitation methods in the literature for particular models include, 
for example, \shortciteN{kadane+dwsp80} and \shortciteN{garthwaite+d88} for linear models, 
and \shortciteN{bedrick+cj96} for generalized linear models, among many others.  
Another popular method for informative prior choice in this tradition is the ``power prior" approach of \shortciteN{chen+i00},
where a tempered version of the likelihood for data from a past study is used as the basis for
the prior;  if no past study is available the data can also be imaginary data created by an expert.  
Extensions or modifications of the  method include \shortciteN{neuenschwander+bs09} and the commensurate priors of \shortciteN{hobbs+cms11}.
However, as mentioned above, we do not focus here on best ways to elicit prior information for particular models, either predictively or on the parameters directly.  
Rather, we are concerned with algorithms for finding good priors satisfying stated prior predictive
constraints already given and where the relevant prior predictive distributions are analytically intractable.

A simple expository example illustrates the main features of our approach.  Suppose we are to observe a binomial
random variable $y\sim \mbox{Binomial}(n,p)$ and we are interested in inference about $p\in (0,1)$.  We parametrize the model
in terms of $\beta=\log (p/(1-p))$ and decide to choose a normal family for the prior on $\beta$, $N(0,\sigma_\beta^2)$, where $\sigma_\beta^2$ is to be chosen.  
We can think of the binomial model with this parametrization as a logistic regression with only an intercept.  A less trivial logistic regression example is 
developed in Section 5.1.  Na\"{i}vely it might be expected that setting $\sigma_\beta^2$ large would result in a non-informative prior. 
However, this is not the case as this would put most of the prior mass far away from $0$ which correspond to values of $p$ near $0$
and $1$.  Setting $\sigma_\beta^2$ small, on the other hand, results in most of the prior mass for $\beta$ near $0$, which corresponds to $p=0.5$.  
So both a large value of $\sigma_\beta^2$, as well as a small value, would usually not be suitable as a non-informative
choice of the prior distribution -- the choice of $\sigma_\beta^2$ requires thought and this example shows that a flat prior that ignores the parametrization
of the model is unacceptable as a non-informative choice.   It is also clear that when $n$ is small, so that there is little information in the data, 
combining what is learned from the data with prior information may be very important, so that 
a non-informative prior choice would not be desirable from that point of view.  
Our logistic regression example in Section 5.1 illustrates the difference that even some limited prior information can make 
to inference in a real example.  While in the case of this example a uniform prior on $p$ may result in inferences with good frequentist properties, 
things become much more complex in multiparameter problems.  
It is well appreciated in the objective Bayesian community that in multiparameter models the specification of a non-informative prior as a reference for an informative
analysis is extremely subtle.  The most successful approach to constructing non-informative priors in a general way is the reference prior approach
(\shortciteNP{berger+bs09}).  However, this approach requires the ability to analytically compute the Fisher information and in general different reference priors
are required for different parameters of interest.  There is simply no such thing as a prior that can be considered non-informative for all functions of the parameter at once. 

There are a variety of ways that prior information is formulated in the elicitation literature.  In our expository example and in view of the
observation that a too diffuse prior would lead to the prior for $p$ concentrating on $0$ or $1$, we might consider the following requirement for the prior.  First, let
$\hat{p}=y/n$ be the maximum likelihood estimator of $p$, and define the summary statistic $S=S(y)=\hat{p}(1-\hat{p})/n$, which
is an estimate of the variance of $\hat{p}$.  
If $p$ is close to $0$ or $1$, we would expect $\hat{p}$ to be close to $0$ or $1$ and $S$ to be small, 
so if the prior predictive for $S$ concentrates on $0$, this indicates the prior is putting most
of its mass near values for $p$ of $0$ or $1$.  
For some suitably chosen small value of $S$, we might require that 
this value be implausible under the prior predictive distribution for $S$ and so rule out such a prior.   
In this simple example it might be more natural to specify prior information on the parameter $p$ directly, but in more complex examples 
prior information may be more easily expressed predictively in terms of observables as we have done here.  
The information we have specified in this case falls short of completely determining a prior, but 
the methods of this paper give ways of exploring prior hyperparameter choices compatible with such information that is easily specified and thought
to be important.  If the analyst feels that the accuracy of any specified prior information is questionable, then, as in any Bayesian analysis, the prior 
should be checked to see if it conflicts with the likelihood as a part of assessing sensitivity of inferences to the prior.

In the next section we describe the basic way that we specify predictive information in the later examples.  We also 
review relevant concepts of Bayesian predictive model checking, since the results of certain model checks for hypothetical data summaries are the way that we formulate
predictive constraints.  Section 3 gives a brief introduction to the literature on history matching and regression ABC methods.  Section 4 then 
discusses the new approach using history matching and regression ABC for prior choice.  Section 5 describes some examples and Section 6 concludes.  

\section{Prior information and Bayesian model checks}

Consider, for a parameter of interest $\theta$, a class of prior distributions $p(\theta|\lambda)$ indexed by a hyperparameter
$\lambda\in \Lambda$.  The problem of prior choice is to choose $\lambda$.  
In predictive elicitation the choice will be based on some characteristics of prior predictive distributions of data or summaries of the data; see \shortciteN[p. 4]{kadane+w98} for a
discussion of the distinction between predictive and structural elicitation.  
Here we will describe one useful way of formulating predictive constraints for elicitation purposes, and certainly
there may be others.  The idea is to use the results of model checks for specified hypothetical data as a way of defining what it means 
for a prior elicitation to be good enough.  In a sense, we treat the problem of elicitation as one of model checking (for hypothetical data).

Suppose there are some summary statistics $S^j=S^j(y)$, $j=1,\dots,J$ of the hypothetical data $y$, with density $p(y|\theta)$, 
and that for these summary statistics we are able to say 
for each one whether certain values should be considered plausible or not under the prior if they were to be observed.  
For $S^j$ we have a vector $h^j$ of hypothetical values supplied by an expert, which we partition as $h^j=(h_I^j,h_P^j)$, where
$h_I^j$ is a vector of values considered as implausible by the expert, and $h_P^j$ is a vector of values considered to be plausible.  
We write $B_I^j$ for the length of $h_I^j$, $B_P^j$ for the length of $h_P^j$, $B^j=B_I^j+B_P^j$ and $B=\sum_{j=1}^k B^j$ for the total
number of constraints.  

In the expository example of the introduction, we considered a $\mbox{Binomial}(n,p)$ model 
parametrized through $\beta=\log (p/(1-p))$ and $\beta\sim N(0,\sigma_\beta^2)$.  Our suggested summary statistic for the elicitation was
the estimated variance of the MLE, $\hat{p}(1-\hat{p})/n$ where $\hat{p}=y/n$, and 
a prior predictive distribution concentrated near zero would indicate an inappropriately large value for $\sigma_\beta^2$ as this 
corresponds to most of the prior mass on $p$ being near $0$ or $1$.  A suitably small implausible value
for the summary here could be obtained by determining a quantile of the summary statistic when the true $p$ is close to $0$ or $1$, say $0.01$ or $0.99$.  

We need to be precise about what plausible and implausible is.  The meaning of these terms will be 
in terms of the result of a prior predictive check (\shortciteNP{box80}).  
Let $p(S^j|\lambda)$ be the prior predictive distribution for $S^j$ under the prior $p(\theta|\lambda)$, i.e.\
$$p(S^j|\lambda)=\int p(S^j|\theta)p(\theta|\lambda)\,d\theta.$$
In the definition, the parameter $\theta$ in the 
sampling distribution for $S^j$ given $\theta$ is integrated out
according to the prior $p(\theta|\lambda)$.  
The prior predictive $p(S^j|\lambda)$ describes beliefs about $S^j$ before any data are observed under the assumed prior $p(\theta|\lambda)$, and is usually not available in
closed form.  
Consider the $p$-values
\begin{align}
 p_{I,b}^j(\lambda) & =P(\log p(S^j|\lambda)\leq \log p(h^j_{I,b}|\lambda)),  \label{pcheck1}
\end{align}
for $S^j\sim p(S^j|\lambda)$ and $j=1,\dots, J$, $b=1,\dots, B_I^j$ and
\begin{align}
p_{P,b}^j(\lambda) & =P(\log p(S^j|\lambda)\leq \log p(h^j_{P,b}|\lambda)),  \label{pcheck2}
\end{align}
where again $S^j\sim p(S^j|\lambda)$ and $j=1,\dots, J$, $b=1,\dots, B_P^j$.  These $p$-values give a measure of how far out in the
tails of $p(S^j|\lambda)$ the various hypothetical summary values are, and hence how surprising they are.  
The $p$-values (\ref{pcheck1}) and (\ref{pcheck2}) are not easy to calculate, and simulation-based methods for approximating them are considered later.
We define a ``reasonable" prior $p(\theta|\lambda)$ in light of the available prior information 
to be one for which given some appropriate cutoff value $\alpha$, we have
$p_{I,b}^j(\lambda)<\alpha$ for $j=1,\dots,J$, $b=1,\dots, B_I^j$ and $p_{P,b}^j(\lambda)\geq \alpha$, $j=1,\dots,J$, $b=1,\dots, B_P^j$  (i.e.\, the values $S^j=h_{I,b}^j$ result in failing a prior predictive check at the cutoff $\alpha$ 
for $j=1,\dots,J$, $b=1,\dots, B_I^j$ and the values $S^j=h_{P,b}^j$, $j=1,\dots, J$, $b=1,\dots, B_P^j$ do not fail such a check).  
Here $\alpha$ is chosen according to the degree of surprise that is considered relevant for the information we want to
put into the prior.  
It is possible also to use a different cutoff $\alpha$ for different checks (and in fact, when eliciting plausible and implausible summaries from an expert, values of $\alpha$ would
need to be given in order to explain to them what plausible and implausible means).  The passing and failing of certain prior predictive checks for hypothetical data summaries
represent constraints on what we consider a reasonable prior to be, and we wish to develop methods for searching the hyperparameter space to find 
corresponding priors satisfying our constraints.  The summary statistics can either be univariate or multivariate.
However, considering a vector valued $S^j$ is more difficult computationally than considering univariate summaries 
due to the need to estimate the prior predictive density in (\ref{pcheck1}) and (\ref{pcheck2}).  In our later examples we generally choose 
univariate $S^j$.  More comments on this, and a cautionary example, are given in Section 5.2.   Generally we would want to choose the summary statistics $S^j$ 
to be reflecting variation related to the parameter $\theta$.  This suggests making these summaries sufficient statistics, although non-trivial minimal
sufficient statistics do not exist in many problems.  Possible choices of the summaries include indicators for the data $y$ belonging to some set (a suggestion made by 
an anonymous referee), or functions of a point estimator if these are available.  Regarding the choice of the hypothetical values, if both plausible and implausible values are specified for a given summary as a pair to 
convey information about the end point of a plausible range, then making these close together is more constraining.  It is important, however, that the chosen values
do not represent information more precise than an expert actually possesses.  

The $p$-values (\ref{pcheck1}) and (\ref{pcheck2}) are examples of prior predictive $p$-values (\shortciteNP{box80}) 
and such $p$-values have in particular found use in the checking for prior-data conflicts when the summary statistic is a minimal sufficient statistic
(\shortciteNP{evans+m06}) and for giving a precise formulation of the notion of a weakly informative prior (as in \shortciteN{evans+j11}, inspired by earlier
work of \shortciteN{gelman06}).  When expressing prior information in terms of the results of model checks, the distinction between
kinds of checks appropriate for different purposes is related to the choice of summary statistics.  This is discussed further in Section 6.
In the application here to problems of prior choice it is natural for us to focus on prior predictive checking.  However, see also
the discussion papers of \shortciteN{gelman+ms96} and \shortciteN{bayarri+b00} or Chapter 5 of \shortciteN{evans15} 
for a variety of perspectives on the broader problem of Bayesian model checking and different types of model checks.  
Now that we have outlined how we specify predictive constraints through prior predictive checks, we need effective methods to
search the space of possible priors.  Our approach adapts the technique of history matching for computer models for this task and this
is described next.

\section{History matching and regression ABC methods:  An overview}

\subsection{History matching}

History matching (\shortciteNP{craig+gss97}) is a method used in the literature for assessing computer models.  A computer model or simulator is a complex computer code that
takes one or more inputs, which we denote as $\lambda$, and produces a set of outputs
$\eta(\lambda)=(\eta_1(\lambda),\dots,\eta_k(\lambda))^T$.   We are reusing our previous notation for prior hyperparameters deliberately here.  
In a history match there are some observed data $y$,
intended to correspond to the computer model outputs, and a so-called implausibility measure, which measures the degree of mismatch
between the observations and the computer model output.  The implausibility measure may be based on some implicit or explicit model allowing for measurement error, 
ensemble variability (the inherent variability of $\eta(\lambda)$ when run multiple times at the same $\lambda$ when the simulator is stochastic) and 
model discrepancy (a model term which represents beliefs about lack of fit of the simulator when run at its best input values).  
In the case of a computationally expensive model, 
we may also wish to use a flexible interpolator such as a Gaussian process (\shortciteNP{rasmussen+w05}) to interpolate or smooth the model outputs $\eta(\lambda)$ based on simulator
runs at a limited number of inputs to reduce computational demands.  Such a model is called an emulator, and emulation uncertainty at inputs where the computer model has not been run can
also be included within the implausibility measure.  

History matching proceeds in waves, starting with a space-filling design covering the range of model inputs ($\Lambda$), and at each wave comes up with a current
non-implausible region for the inputs, reducing the size of the non-implausible region at each stage.  The phrase non-implausible rather than plausible
is used since the non-implausible region consists only of the region of the space not ruled out yet as unsuitable.
The iterative aspect of the process allows us to place more points adaptively in ``promising" regions of the input space $\Lambda$, something which is important
when $\lambda$ is high dimensional.    If emulation is used for a computationally expensive model, this adaptive aspect, where more model evaluations
are made in the interesting part of the space allows the quality of emulation to improve as more waves are considered.  
Thresholds on the implausibility measure determining the current implausible region 
may become more stringent as the waves proceed and different observations may also be introduced sequentially in this process.
The philosophy of history matching is not to find a ``best input" for the model, but to explore the space of non-implausible values for the model parameters.
The non-implausible region at the end of the process may be empty if there are no parameters providing an adequate
fit to the outputs.  A history match can be instructive for guiding model development, and if a model
is good enough to warrant the computational expense of calibration then the history match can be useful for developing efficient computational algorithms.
History matching has been successfully used in petroleum reservoir modelling (\shortciteNP{craig+gss97}), 
galaxy formation models (\shortciteNP{vernon+gb10,vernon+gb14}), rainfall-runoff models (\shortciteNP{goldstein+sv13}), 
climate models (\shortciteNP{williamson+others13}) and infectious diseases models (\shortciteNP{andrianakis+others15}) among other applications.
Relationships between history matching and approximate Bayesian computation (ABC) algorithms have been considered recently by
\shortciteN{wilkinson14} and \shortciteN{holden+others2015}.

Given an implausibility measure $I(\lambda)$, history matching proceeds in the following way.
\begin{enumerate}
\item Initialization.  Set $w=1$ and generate a collection of $r$ points $\lambda_1^{(1)},\dots,\lambda_r^{(1)}$ for $\lambda$ according to a space-filling
design covering the range of the inputs, $\Lambda$.  
\item Until some stopping rule is satisfied:
\begin{enumerate}
\item Calculate $I(\lambda_1^{(w)}),\dots,I(\lambda_r^{(w)})$.
\item Choose some subset of the collection of the current inputs, $\lambda_1^{(w)},\dots,\lambda_q^{(w)}$, as non-implausible based on thresholding the implausibility measure.  
This set of points is used to define a current non-implausible region $N^w$.  
\item Generate points $\lambda_1^{(w+1)},\dots,\lambda_r^{(w+1)}$ according to a new space-filling design covering $\Lambda^w$ and set $w=w+1$.  
\end{enumerate}
\end{enumerate}
In Section 4 we describe how we implement the steps in the procedure above for our later applications.  There are a variety of approaches in the existing
history matching literature for the construction of the implausibility measure, the construction of space filling designs and other choices.
In different applications the implausibility measure might change between iterations or only a subset of observations might be considered in the
early stages and the implausibility thresholds might change between iterations. 
In our later applications, at wave $w$, the wave $w+1$ samples are generated directly from the current ones without explicitly defining the set $N^w$, and so we don't describe how this set is sometimes constructed in the history matching literature.  A variety of approaches to this issue may be found in the above references.
If an emulator is used in evaluation of the implausibility measure, additional model evaluations could be made at step 2 (b) for the current non-implausible points and the
emulator updated appropriately.  These additional model evaluations and updating of the emulator may be particularly important in the case of high-dimensional models, and
the task of emulation becomes much simpler as the interesting region of the space shrinks over successive waves. See Algorithm 1 of \shortciteN{drovandi+np17} for a typical implementation of
history matching with sequential updating of an emulator.

\subsection{Regression ABC methods}

ABC methods are used in the Bayesian analysis of models where the likelihood is intractable
(\shortciteNP{tavare+bgd97,pritchard+spf99,beaumont+zb02}).  
The basic idea of simple ABC methods is to conduct forward simulations from the model according to parameter
values sampled from the prior and to then see
whether the simulated data are similar to the observed data.  If it is, then the parameter value that generated the simulated data is retained as one that might plausibly
have generated the data.  A recent review of these methods is given by \shortciteN{marin+prr11}, but here we
confine ourselves to describing only some regression based approaches used in the ABC literature which are relevant to the calculations done in the
next section (\shortciteNP{beaumont+zb02,blum+f10}).  

Suppose that $p(\theta|\lambda)$ is the prior, $p(y|\theta)$ is the data model and $y_{obs}$ is the observed data.  In ABC one simulates $(\theta_i,y_i)$, $i=1,\dots,I$ 
from the prior and then the simulated data are reduced to a summary statistic $S_i=S(y_i)$ with $S_{obs}=S(y_{obs})$.  The role of summary statistics in
an ABC analysis is to reduce the dimensionality of the data, and ideally the summary statistics should be nearly sufficient for $\theta$.  
The idea of regression based ABC methods is to use regression to obtain a conditional density estimate
of $\theta$ given $S_{obs}$ (i.e.\ to approximate the posterior distribution $p(\theta|S_{obs})$).  We assume that $S_{obs}$ contains most of the relevant
information about $\theta$ in $y_{obs}$.  \shortciteN{blum+f10}, extending 
methods originally due to \shortciteN{beaumont+zb02}, consider the regression
model 
\begin{align}
 \theta_i & =\mu(S_i)+\sigma(S_i)\epsilon_i,  \label{regmodel1}
\end{align}
where $\mu(\cdot)$ and $\sigma(\cdot)$ are flexible mean and standard deviation functions (which they parametrize using neural networks) and the $\epsilon_i$
are zero mean variance one residuals.  It is assumed above that $\theta$ is a scalar parameter, but extensions to the multivariate case are straightforward in which
$\mu(S)$ and the $\epsilon_i$ are multivariate and $\sigma(S)$ is a matrix square root of the covariance matrix of $\theta$ given $S$.  To obtain an approximate 
sample from $\theta|S_{obs}$, which we write as $\theta_i^a$, $i=1,\dots,I$  (i.e.\ an approximate sample from the posterior) we can consider fitting the regression model
to obtain estimates $\hat{\mu}(\cdot)$ and $\hat{\sigma}(\cdot)$ of $\mu(\cdot)$ and $\sigma(\cdot)$ respectively, and then use empirical residuals in the fitted
regression at $S=S_{obs}$:
$$\theta_i^a=\hat{\mu}(S_{obs})+\hat{\sigma}(S_{obs})\hat{\epsilon}_i=\hat{\mu}(S_{obs})+\hat{\sigma}(S_{obs})\hat{\sigma}(S_i)^{-1}(S_i-\hat{\mu}(S_i)),$$
$i=1,\dots,I$.  In the discussion above it is also possible to localize the regression using a kernel function and attach weights to the adjusted sample values 
$\theta_i^a$ (\shortciteNP{blum+f10}).  

\shortciteN{nott+dme15} consider related methods for repeated
conditional density estimation when we want to simulate from a data model for different values of a parameter and where that is expensive.  
For approximate simulation from the data model
the roles of $S$ and $\theta$ are reversed in (\ref{regmodel1}).  That is, we consider
\begin{align}
 S_i & =\mu(\theta_i)+\sigma(\theta_i)\epsilon_i,  \label{regmodel2}
\end{align}
and then for a given $\theta$ an approximate sample from $S$ given $\theta$ would be
$$S_i^a=\hat{\mu}(\theta)+\hat{\sigma}(\theta)\hat{\sigma}(\theta_i)^{-1}(\theta_i-\hat{\mu}(\theta_i)),$$
for estimates $\hat{\mu}(\theta)$ and $\hat{\sigma}(\theta)$ of $\mu(\theta)$ and $\sigma(\theta)$.  In the next
section we use a model similar to (\ref{regmodel2}) to simulate in a computationally thrifty way from a prior predictive distribution
$p(S|\lambda)$ for summary statistics $S$ conditional on a prior hyperparameter $\lambda$ with $\theta$ integrated out according to the prior $p(\theta|\lambda)$.  
Such approximate prior predictive samples are useful for estimating $p(S^j|\lambda)$ (a quantity which appears in our prior predictive $p$-values
(\ref{pcheck1}) and (\ref{pcheck2})) and hence for choosing an appropriate value of $\lambda$.   

\section{Proposed algorithm for prior choice}

Our proposed algorithm applying history matching for prior choice will now be described.  Let $\lambda$ denote the prior hyperparameters in a problem of prior choice.  Given $\lambda$
we can compute certain features of prior predictive distributions as outputs of the Bayesian model.  In the procedure of Section 2 we may consider the outputs to
be the $p$-values in equations (\ref{pcheck1}) and (\ref{pcheck2}).  From these an implausibility measure can be constructed based on desired constraints for the outputs.  Later we use
the implausibility measure 
\begin{align}
 I(\lambda) & = \sum_{j=1}^J \sum_{b=1}^{B^j_I} \max (0,p_{I,b}^j(\lambda)-\alpha)+\sum_{j=1}^J\sum_{b=1}^{B_P^j} \max (0,\alpha-p_{P,b}^j(\lambda)) \label{implausibility}
\end{align}
and we note that $I(\lambda)$ is $0$ if the constraints considered in Section 2 are satisfied, i.e.\ $p_{I,b}^j(\lambda)<\alpha$, $j=1,\dots,J$, 
$b=1,\dots,B_I^j$ and
$p_{P,b}^j(\lambda)\geq \alpha$, $j=1,\dots,J$, $b=1,\dots, B_P^j$, with $I(\lambda)>0$ if one or more of these constraints are violated.  

Consider once more the expository example of the introduction.  There we considered for the binomial model $\mbox{Binomial}(n,p)$ parametrized
by $\beta=\log (p/(1-p))$ the summary statistic $\hat{p}(1-\hat{p})/n$ with $\hat{p}=y/n$, and suggested defining some small
value of this statistic as implausible as a way of constraining the prior to not place too much mass near values for $p$ of $0$ or $1$.  
In this example there is just a single $p$-value, corresponding to an implausible summary, and the above implausibility measure 
is given by this $p$-value minus $\alpha$ if the $p$-value is bigger than $\alpha$, and zero otherwise.  

The search for prior hyperparameters satisfying the constraints can be performed using the methods of history matching with the implausibility measure (\ref{implausibility}).  
One might object that the threshold $\alpha$ used in our implausibility is somewhat artificial.  However it should be kept
in mind that this threshold is not used in a binary decision making context here, and that the purpose of $I(\lambda)$ is just to guide the search
to a fruitful region of the hyperparameter space.  Obtaining an exactly $0$ value of $I(\lambda)$ may not be so important.  
The use of $p$-values in $I(\lambda)$ is convenient for the way that it puts information from the different summary statistics on the same scale, and 
we have found the choice (\ref{implausibility}) for the implausibility measure to be useful although there are certainly other ways that the implausibility
could be defined.  

Steps 2 b) and c) of the history matching algorithm given in Section 3.1 for wave $w$ are implemented in our later examples in the following way.  
First, choose some fraction $\gamma$ of $r$ in such a way
that both $1/\gamma$ and $Q=\gamma r$ are integers. For instance, in the first example of Section 5
we use $\gamma=0.1$ and $r=100$.  Next, choose the $Q$ values of $\lambda$ in the current
wave for which $I(\lambda)$ is smallest.  Write these values as $\lambda_1^{* (w)},\dots, \lambda_Q^{* (w)}$.  Then for each of $q=1,\dots,Q$, generate
$1/\gamma$ values from a normal distribution $N(\lambda_k^{* (w)},\Sigma^{(w)})$ where $\Sigma^{(w)}=h^2 V_w$, $V_w$ is the sample covariance matrix
of all the wave $w$ samples, and $h=\left(\frac{4}{(2d+1)Q}\right)^{1/(d+4)}$ where $d$ is the dimension of $\lambda$.  Note that
this results in $Q/\gamma=r$ samples that we take as the wave $w+1$ samples.  
In our later examples we use the modified sampling approach in the {\tt mvrnorm} function in the R package {\tt MASS} (\shortciteNP{venables+r12})
with the option {\tt empirical=TRUE} 
to obtain generated samples that have exactly the sample covariance matrix $\Sigma^{(w)}$.  
The definition of $\Sigma^{(w)}$ in the
sample generation step is obtained by inflating an automatic choice of kernel bandwidth used in the multivariate kernel density estimation literature by a factor of $4$ 
(\shortciteNP{silverman86}). 
There are other ways to generate a space-filling design for each wave;  the idea above and that we implement
later in examples is a simple one based on a similar suggestion in Andrianakis et al. (2015) based on perturbing values according to a normal kernel with enough variability to
ensure that the new points are sufficiently different to the current one.  
The intuition behind our choice for $h$ is that after pruning away the implausible samples in the current wave, we want to generate a set of points for the next wave that covers the distribution for the current set of non-implausible samples.   The kernel estimate with bandwidth choice given above
is just to make the next wave samples somewhat overdispersed compared to the distribution of current non-implausible samples.  
Note that if we were to simulate from the kernel density estimate fitted to the current non-implausible samples, that would correspond to choosing one of the non-implausible
samples at random and then drawing from a normal density centered on that sample.  Instead of choosing a point randomly in this process, if we ensure all the non-implausible
samples are represented equally when drawing the next wave samples, we arrive at the procedure we have suggested above.
Inflating the bandwidth choice of \shortciteN{silverman86} by 4 doubles the marginal standard 
deviations used in local perturbations of the current samples in the process of simulating the next wave samples.
It is difficult to say anything about optimality of our suggested choice of $h$.  
A larger value of $h$ will ensure that the non-implausible region is not collapsed down too quickly, at the expense of additional computations.  
How quickly we should narrow down the non-implausible region also interacts with how many samples are used in the initial space-filling design, and how smooth
the implausibility measure is.
The only remaining detail to specify in the algorithm is the stopping rule.  A useful stopping rule is to stop when either a zero implausibility value has been found, 
or if there has been no further decrease in the minimum implausibility value found for a certain number of waves.  

Computing the implausibility measures in the application of history matching to prior choice as discussed in Section 3 involves computation
of the $p$-values in equations (\ref{pcheck1}) and (\ref{pcheck2}) for a large number of different values of $\lambda$ and this can be computationally burdensome.  Our solution is
to use the regression approximate Bayesian computation (ABC) methods introduced in Section 3.2 to approximate these $p$-values in a computationally thrifty way.
The methods considered are based on those developed in \shortciteN{nott+dme15}, and 
play a similar role in our later examples to the role of emulators in history matching for computationally expensive computer models.  

Suppose we wish to approximate $I(\lambda)$ for a possibly large set of different $\lambda$ values, 
$\lambda^n$, $n=1,\dots,N$.  These values might be a grid over the region of interest for $\lambda$ if $\lambda$ is low-dimensional, or in the history matching
procedure they might be the hyperparameter values generated in the current wave. 
Let $p(\lambda)$ be a pseudo-prior for $\lambda$ which covers the range of the values of $\lambda$ of interest.  This pseudo-prior is not to be used for inference
but is used in generation of samples of the summaries $S^j$.  We simulate values 
$(\lambda_i,\theta_i,y_i)$ from $p(\lambda)p(\theta|\lambda)p(y|\theta)$, $i=1,\dots,I$ independently.  From the $y_i$ we obtain simulated
summaries $S^j_i=S^j(y_i)$, $i=1,\dots,I$, $j=1,\dots,J$.  We can obtain an approximate sample from $p(S^j|\lambda)$ for any given value of 
$\lambda$ by considering the regression adjustment methods of Section 3 applied to the regression model
$$S_i^j=\mu^j(\lambda_i)+\sigma^j(\lambda_i)\epsilon_i,$$
where the $\epsilon_i$ are independent and identically distributed errors with mean zero and variance one and $\mu^j(\lambda)$ and $\sigma^j(\lambda)$ are
flexible mean and standard deviation functions.  This is similar to the regression adjustment approach considered for equation (\ref{regmodel2}) in Section 3 applied to the marginalized
model for the summaries where $\theta$ has been integrated out according to $p(\theta|\lambda)$.  
Extension to the case where $S_i^j$ is multivariate can
also be considered but in our later examples the $S^j$ are each univariate summaries. 
Fitting the regression model locally, based on a certain number of nearest
neighbours of $\lambda$, is often useful.  This is something we consider later in the examples with a nearest
neighbour distance following the default choice in the {\tt R} package {\tt abc} (\shortciteNP{csillery+fb12}).  Although we do not describe in detail
the implementation of regression adjustment in the {\tt abc} package, for the method of \shortciteN{blum+f10} $\mu^j(\cdot)$ and $\sigma^j(\cdot)$ 
are parametrized by neural network models, and these functions are estimated in a two step procedure.  In the first step, the mean function is estimated assuming
the variance is constant.  Then the logarithm of the variance function is estimated by fitting a second neural network model to the logarithm of the squared residuals 
from the first stage fit.  The fitting can be localized, in the sense that only a certain number of nearest neighbour points closest to the target covariate value are used 
(where closest is in the sense of a scaled
Euclidean distance, with the scaling for each covariate based on the mean absolute deviation of values for the covariate).  
The {\tt abc} package also implements linear regression (\shortciteNP{beaumont+zb02}) and other regression adjustments.  In general, there can be a trade-off between
the flexibility of the regression model used for the adjustment, and the size of the neighbourhood required with less flexible regression models
requiring smaller neighbourhoods.  
As mentioned above we use the default
tuning parameter values implemented in the {\tt abc} package and refer the reader to \shortciteN{csillery+fb12} for further details.  

An approximate sample from $p(S^j|\lambda^n)$ is
\begin{align}
 \hat{S}_i^{j,n}= & \hat{\mu}^j(\lambda^n)+\hat{\sigma}^j(\lambda^n)\hat{\sigma}^j(\lambda_i)^{-1}(S_i^j-\hat{\mu}^j(\lambda_i)),\;\;\;  i=1,\dots,I, \label{ppsamples}
\end{align}
and then we can construct a kernel density estimate of $p(S^j|\lambda^l)$, written $\hat{p}(S^j|\lambda^l)$, from these approximate samples.  
The kernel density estimate is constructed independently for each summary statistic.  
How close this kernel density estimate is to the predictive density it approximates depends on how well the regression adjusted samples approximate a draw from the predictive
density, as well as other factors such as the kernel, sample size and bandwidth choice.  
The quality of the regression adjusted samples for approximating a sample from the true prior predictive can be very good
if the regression fitting is done in a small neighbourhood and that neighbourhood contains a large number of samples.  If the predictive density varies smoothly with
$\lambda$  then the predictive density changes very little throughout a small neighbourhood of the targeted $\lambda$ value.  When fitting locally with sufficient
samples the regression adjustment has little effect and the regression adjusted sample is indistinguishable from a sample from the true prior predictive as the neighbourhood
shrinks.   Of course, achieving a very small neighbourhood size containing a large number of samples in local fitting involves simulating a large number 
of summary statistic values and a heavy computational burden.

The computation of the estimated $p$-values $\hat{p}_{I,b}^j(\lambda^n)$, $j=1,\dots, J$, $b=1,\dots, B_I^j$ and $\hat{p}_{P,b}^j$, $j=1,\dots, J$, $b=1,\dots, B_P^j$, can be performed using the following algorithm.
\begin{enumerate}
\item Given the input hyperparameter $\lambda^n$, obtain approximate samples $\hat{S}_i^{j,n}$, $i=1,\dots,I$ from $p(S^j|\lambda^n)$, $j=1,\dots,J$, according to
(\ref{ppsamples}).
\item For each statistic $S^j$, $j=1,\dots, J$, calculate a kernel estimate of $p(S^j|\lambda^n)$ at $\hat{S}_i^{j,n}$, $\hat{p}(\hat{S}_i^{j,n}|\lambda^n)$, $i=1,\dots,I$, $h_{I,b}^j$, $b=1,\dots, B_I^j$ and $h_{P,b}^j$, $b=1,\dots, B_P^j$.
\item Calculate
$$\hat{p}_{I,b}^j(\lambda^n)=\frac{1}{I}\sum_{i=1}^I I(\log \hat{p}(\hat{S}_i^{j,n}|\lambda^n)\leq \log \hat{p}(h_{I,b}^j|\lambda^n)),\;\;
j=1,\dots, J,\;\; b=1,\dots, B_I^j,$$
and
$$\hat{p}_{P,b}^j(\lambda^n)=\frac{1}{I}\sum_{i=1}^I I(\log \hat{p}(\hat{S}_i^{j,l}|\lambda^n)\leq \log \hat{p}(h_{P,b}^j|\lambda^n)),\;\;
j=1,\dots, J, \;\;b=1,\dots, B_P^j.$$
\end{enumerate}
Given the estimated $p$-values for a certain $\lambda^n$ we can check whether it is acceptable according to our criteria by checking if
$\hat{p}_{I,b}^j(\lambda^n)<\alpha$, $j=1,\dots,J$, $b=1,\dots,B_I^j$ and $\hat{p}_j(\lambda^n)\geq \alpha$, $j=1\dots,J$, $b=1,\dots, B_P^j$. 
An approximate implausibility value $I(\lambda^n)$  can also be computed from the $p$-values.
Note that the regression ABC computations are being used in a
screening process to remove highly implausible values of the hyperparameters
and high precision is not needed.
Once a hyperparameter value is chosen
based on the regression calculations as giving a prior satisfying the desired constraints
we can check its suitability.  
We can do this by generating a large number of values of $S^j$, $j=1,\dots,J$ from the prior predictive distribution for
the chosen $\lambda$, and from these approximate the $p$-values accurately,
to check that
the regression approximations were good enough.  
Such a procedure would not be feasible for a large number of different candidate values of $\lambda$, 
which is why the regression approximations are used within the history matching algorithm.  However, after the history matching is completed and
we have identified one or a small number of suitable $\lambda$, it is quite feasible to
generate a large sample from the prior predictive distribution for these, without using the regression methods, in order to confirm their suitability.
 
The approach we have described of approximating prior predictive samples based on 
local regression adjustments can fail when the prior predictive density changes rapidly as a function of $\lambda$, and it may also be difficult to apply in high dimensions.  
It is also assumed above
that summary statistics are generated once at the beginning of the history match according to values for $\lambda$ simulated under the pseudo-prior $p(\lambda)$.  It was mentioned in Section 3 that a powerful aspect of history matching is the way that additional model evaluations (or summary statistic simulations in the present case) can be made as the waves of the history matching proceed.  That is, we can generate additional summary statistic simulations at each of the current non-implausible $\lambda$ values in the history matching waves to improve the quality of the regression adjustment approach for approximating the prior predictive distribution in the interesting parts of the hyperparameter space.  This is most interesting when the number of hyperparameters is large, and for our highest dimensional example later (with four hyperparameters) we consider such an approach.
Emulation methods are thoroughly developed in the existing literature for deterministic computer models.
However, where stochastic models are considered, and the task is to emulate the distribution of an output
as a function of inputs, simple methods such as just emulating means and variances are often considered.  This may be sufficient, depending on what is required for the chosen implausibility measure.  In our application, capturing more complex features of the prior predictive density becomes important.  The regression ABC approach outlined here
is not the only one that could be considered.  However, a comparison of different
conditional density estimation methods in this application is beyond the scope of the present work.  

\section{Examples}

We illustrate our methodology in three examples.  In the first two examples there are just two hyperparameters to be chosen and we can plot the way that
the predictive $p$-values in our checks vary with the hyperparameters over a grid;  such plots are useful for checking the results of the history match.  
Both the $p$-values at the grid points in these plots, as well as the $p$-values used to approximate the implausibility measure for the history matching samples, 
are obtained using regression ABC approximations to the prior predictive densities of the summaries.  
In the third example there are four hyperparameters to be chosen, and consideration of a grid of hyperparameter values is no longer feasible.  

\subsection{Logistic regression example}

We consider a logistic regression for an experiment described in \shortciteN{racine+gfs86} where 5 animals at each of 4 dose levels
were exposed to a toxin.  We write the dose levels as $x_1<x_2<x_3<x_4$ and assume that these values have been transformed to a log scale, centered and scaled 
as in \shortciteN{gelman+jps08}.  If $y_i$ is the number of animals killed at dose level $x_i$, the data model is $y_i\sim \mbox{Binomial}(5,p_i)$ with
$\log (p_i/(1-p_i))=\beta_0+\beta_1 x_i$.  \shortciteN{gelman+jps08} consider a prior on $\beta$ where $\beta_0$ and $\beta_1$ follow independent Cauchy distributions
centered on zero with scale $\lambda_1=10$ and $\lambda_2=2.5$ respectively.  Here we consider $\lambda=(\lambda_1,\lambda_2)$ as hyperparameters to be chosen, with
$\lambda\in [0.5,10]\times [0.5,10]$.  

Our elicitation method requires us to specify some hypothetical data to be plausible or implausible under the prior.  Write $\hat{\beta}=(\hat{\beta}_0,\hat{\beta}_1)$ 
for the posterior mode of $\beta$
based on independent normal $N(0,100)$ priors on $\beta_0,\beta_1$.    Note that the normal prior here is used only in the computation of $\hat{\beta}$:  the 
parametric prior family being used in the elicitation is the Cauchy family described above.
Note that $\hat{\beta}$ is similar to the MLE in non-degenerate settings but will exist even when the MLE does not.  
For each dose $x_i$, let $\hat{p}_i=1/(1+\exp(-\hat{\beta}_0-\hat{\beta}_1 x_i))$ be
the corresponding fitted probability of death at dose $x_i$ under the fitted model.  Let us consider the summary statistic $S^1=\sum_{i=1}^4 5\hat{p}_i(1-\hat{p}_i)$
which is the sum of the variances of the responses when $\beta=\hat{\beta}$.  The statistic $S^1$ will tend to be small if all the responses are close
to either zero or the maximum value of $5$ resulting in fitted probabilities at the different dose levels all close to zero or one.  If all $\hat{p}_i$ are equal to either $0.01$ or $0.99$, then the value of $S^1$ would be $0.198$ and we might wish the prior to express the information that this is an implausible value for $S^1$.  
The summary $S^1$ is the natural extension to the logistic regression case of the summary statistic used in the expository example of the introduction.  

In this example we might also expect that it would not be surprising if the fitted probability of death goes from a value near zero at the lowest dose to a value near 1 at the highest dose, in a fairly smooth way.  
If $\hat{p}_1=0.01$, $\hat{p}_2=0.25$, $\hat{p}_3=0.75$ and $\hat{p}_4=0.99$, then the corresponding value of $S^1$ would
be $1.974$.  We consider a prior within our framework in which  $S^1=0.198$ is considered to be implausible, and $S^1=1.974$ is considered
to be plausible.  This is weak prior information, but enough to constrain hyperparameter 
choice in a useful way.  Although it is discrete, $S^1$ is treated as a continuous quantity in our calculations.  This is a reasonable approximation
when the number of different possible values is large, as here.

\begin{figure}
\begin{center}
\begin{tabular}{cc}
\includegraphics[width=70mm]{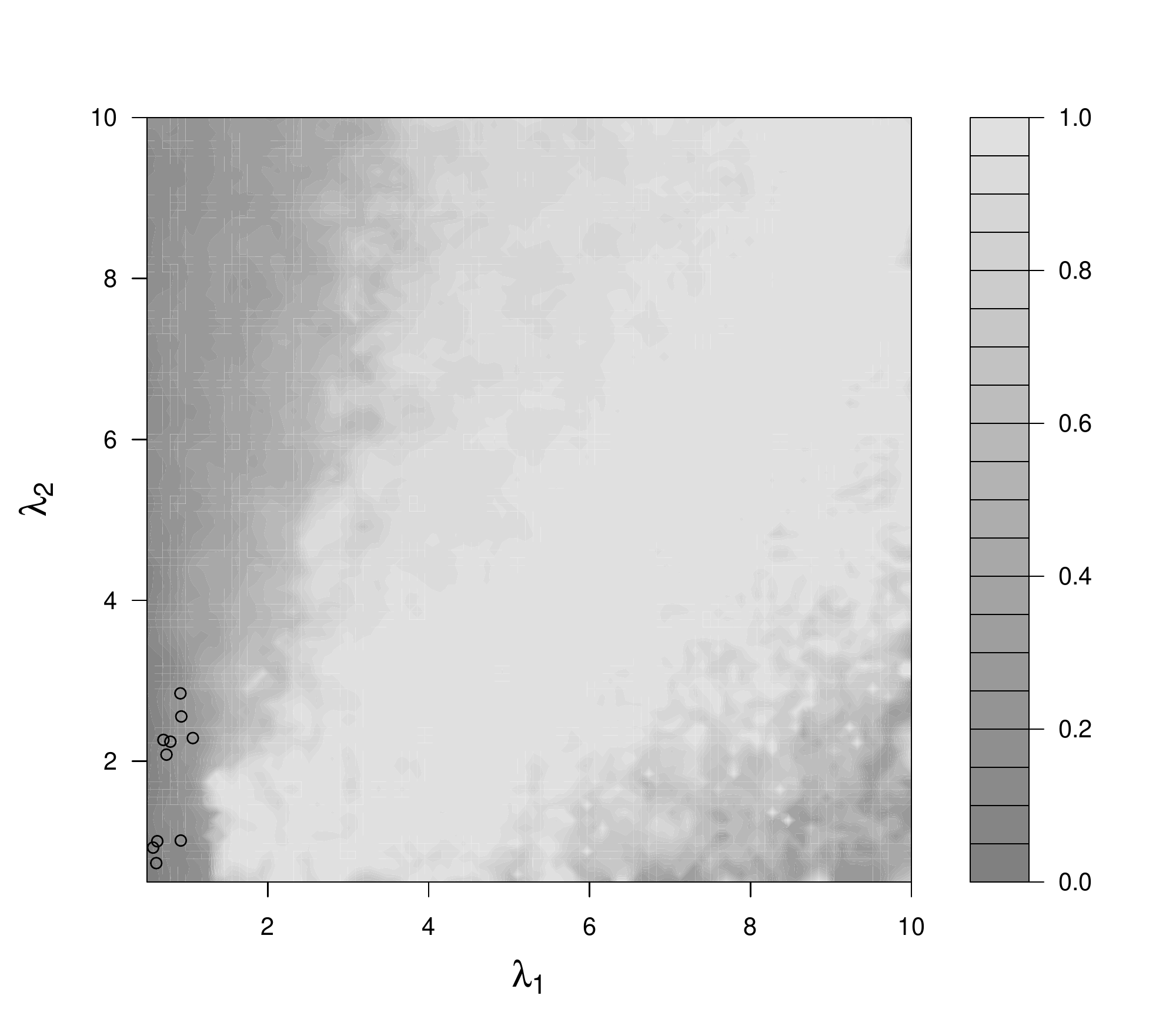}  & 
\includegraphics[width=70mm]{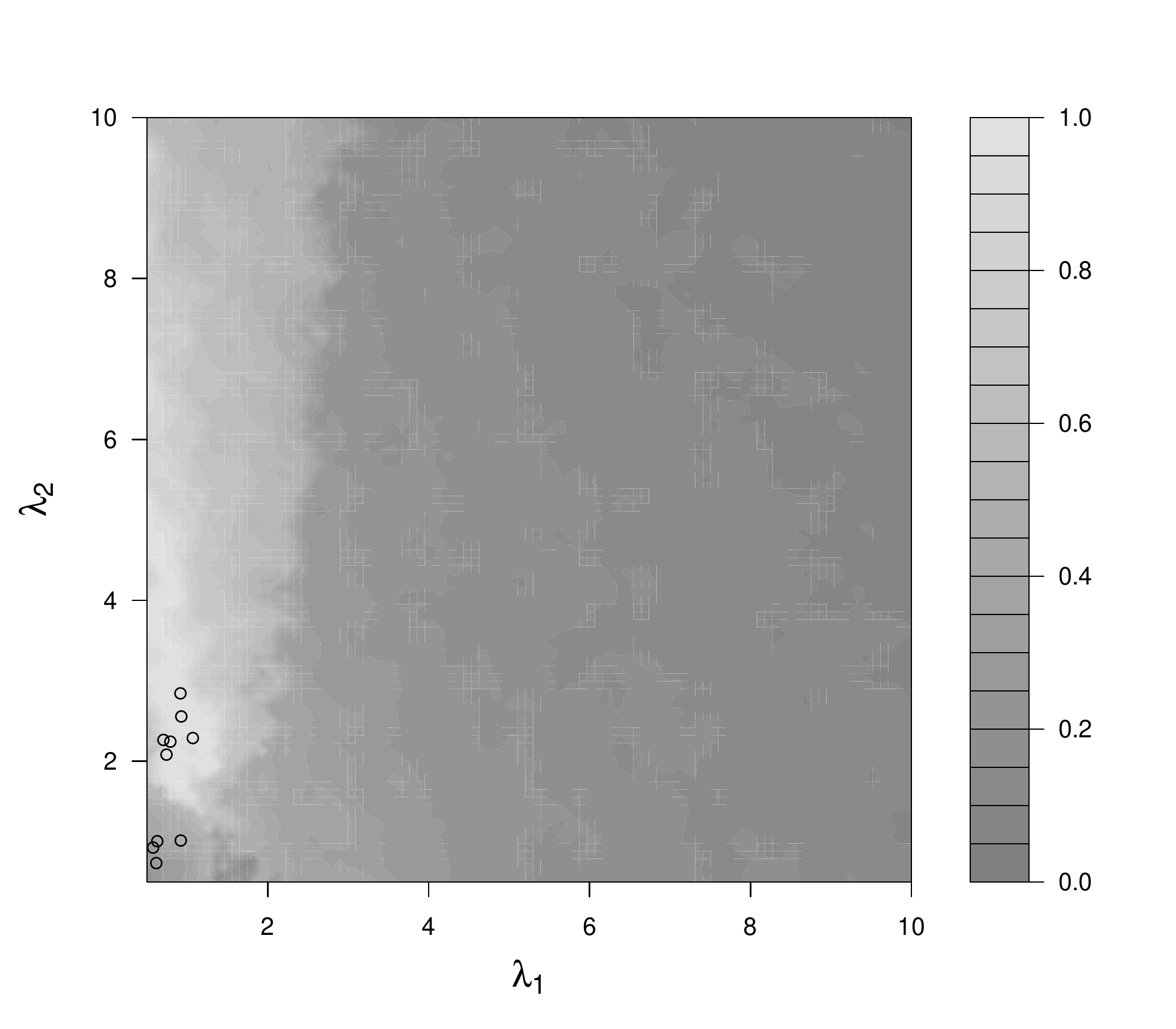}
\end{tabular}
\end{center}
\caption{\label{pvalueplot} 
Conflict $p$-value as a function of $\lambda$ for logistic regression example.  $p$-value for check for $S^1=0.198$ (left) and for $S^1=1.974$ (right).  
In both graphs the overlaid points are from the fourth wave of the history match and the minimum implausibility obtained is zero.
}
\end{figure}

\begin{figure}
\begin{center}
\begin{tabular}{cc}
\includegraphics[width=70mm]{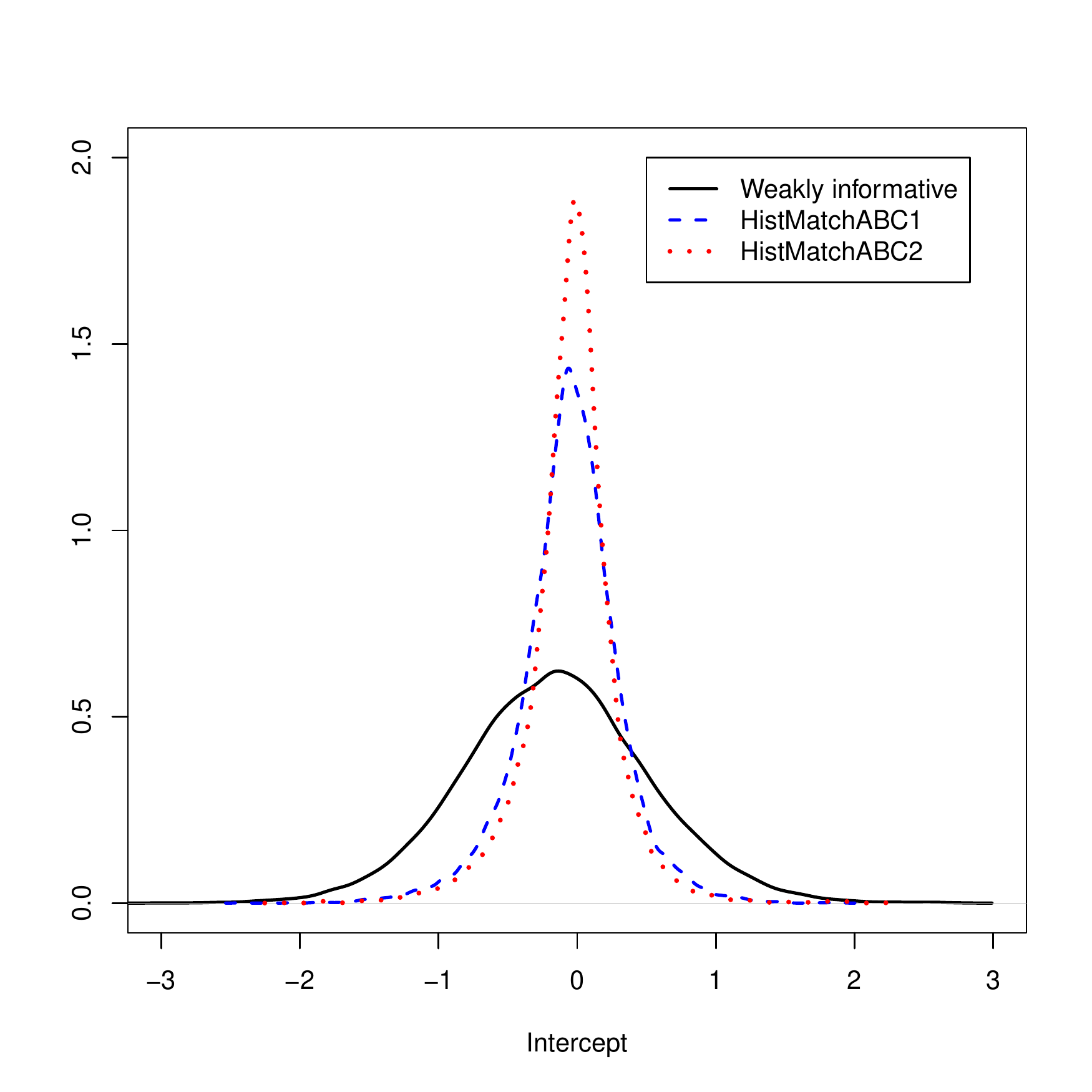}  & 
\includegraphics[width=70mm]{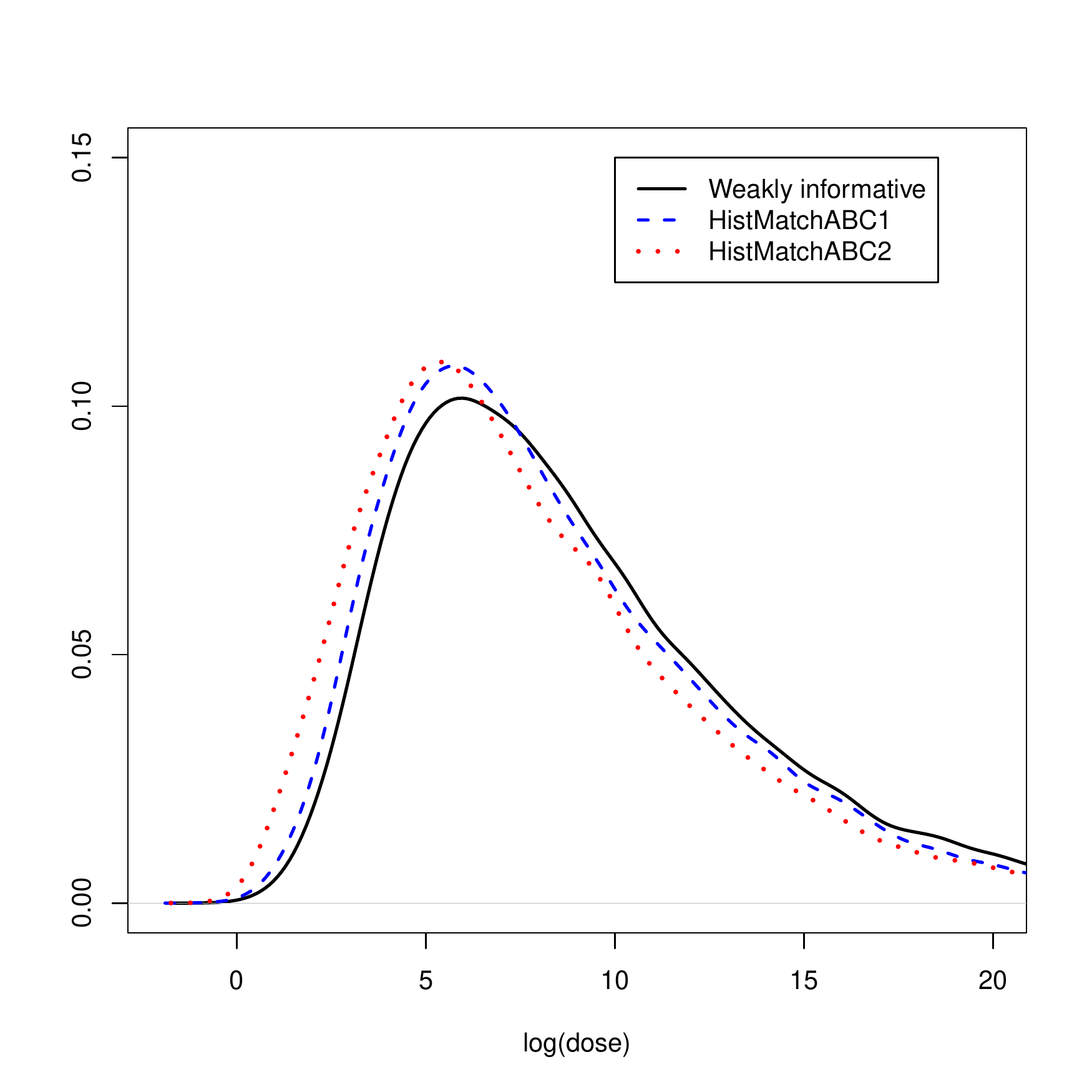}
\end{tabular}
\end{center}
\caption{\label{whatdiff} 
Marginal posterior distributions for $\beta_0$ (left) and $\beta_1$ (right) for default prior with $(\lambda_1,\lambda_2)=(10,2.5)$ as well as  history matching hyperparameter values of 
$(\lambda_1, \lambda_2) = (0.33, 2.08)$ and $(\lambda_1, \lambda_2) = (0.23, 0.73)$ (labelled ``HistMatchABC1" and ``HistMatchABC2" respectively).}
\end{figure}

For the hypothetical data summary $S^1=0.198$, we compute the predictive $p$-value for the summary statistics chosen using the method of Section 4 and using
 a grid of 10,000 $\lambda$ values in our target range
$\lambda\in [0.5,10]\times [0.5,10]$ with the grid formed from 100 equally spaced values in each dimension.  
The regression adjustment calculations for computation of the $p$-values are done using the default implementation of the {\tt abc} function in the {\tt abc} {\tt R} package 
(\shortciteNP{csillery+fb12}).  We used 400,000 simulated values of the summary statistic $S^1$, local
linear regression adjustment and 1,000 nearest neighbours in the localized regression ABC procedure. 
This means that in (\ref{ppsamples})  the mean and log standard deviation functions $\mu^j(\lambda)$ and $\log \sigma^j(\lambda)$ are assumed to be linear functions
of $\lambda$, and the regression is fitted based on the nearest 1000 neighbours to the target $\lambda$ values.  Nearest means in the sense of scaled Euclidean distance, 
where each component of $\lambda$ is being scaled by the mean absolute deviation.  This is the default local linear regression adjustment implemented 
in the {\tt abc} R package (\shortciteNP{csillery+fb12}).  
A plot of how the $p$-value changes
as a function of $\lambda$ is shown in the left panel of Figure \ref{pvalueplot}.  Note the two blue regions in the graph where the $p$-value is small;  the region on the left occurs
for hyperparameter values where $0.198$ is an implausibly small value, whereas the region on the right occurs for hyperparameter values for which $0.198$ is implausibly large.  
A similar plot of the $p$-value as a function of $\lambda$ for the check with $S^1=1.974$ is
shown in the right panel.  An acceptable value for $\lambda$ is a value in the dark grey region in the left panel (small $p$-value indicating a prior-data conflict) and avoiding the dark 
grey region in the right panel (a $p$-value which is not small indicating the absence of a conflict).    
The points overlaid on the graphs are obtained from using the history matching method of Section 3.  In the history match the algorithm is initialized with a maximin latin hypercube
design of $r=100$ points, $\gamma=0.1$ and the points shown in the graph are the retained values after 4 waves.  The $p$-values in the implausibility measure
are again computed using the method of Section 4.  The minimum implausibility obtained is $0$, i.e.\ we are successful at finding hyperparameter values satisfying the constraints.  
As mentioned above, in considering this example \shortciteN{gelman+jps08} considered a default prior with $\lambda_1=10$ and $\lambda_2=2.5$.  This is a weakly informative 
choice for the prior, and it
can be seen from Figure \ref{pvalueplot} that to match the information we have suggested putting into our analysis a smaller value of $\lambda_1$ is needed.   
Also shown in Figure \ref{whatdiff} are the marginal posterior distributions of $\beta_0$ and $\beta_1$ for the default prior with $\lambda_1=10$ and $\lambda_2=2.5$, as well as for two hyperparameter values obtained from the history match.  The posterior distributions are computed for the observed data of $(y_1,y_2,y_3,y_4)=(0,1,3,5)$.  In this example it is seen that the prior information we have put in makes some difference to the resulting inference, particularly for the intercept.

\subsection{Sparse signal shrinkage prior}

Next we consider prior choice for a linear model with a sparse signal shrinkage prior on the coefficients.  The shrinkage prior we consider
is the horseshoe+ prior of \shortciteN{bhadra+others2015}.  
The need in modern data analysis to consider increasingly complex models with respect to both the number of parameters and hierarchical structure 
has resulted in a very large literature on sophisticated shrinkage priors in a range of applications.  We consider only the horseshoe+ prior for a high-dimensional linear model
in this example, but the kind of analysis we do here could be done for other shrinkage priors, of which there are many.  
\shortciteN{bhadra+others2015} give a survey of the current state of the art in the area.
We describe a general version of our model first which also incorporates
observation specific mean shift terms that can account for outliers in the model, using similar ideas to those considered in \shortciteN{she+o11}.
A simplified version of the model with two hyperparameters will be considered in this subsection, and the more general form of the model with four hyperparameters
will be considered in the next subsection. 

For some $(M\times p)$ design matrix $X$ consider the model
\begin{align}
y & =\beta_0 1_M+X\beta+\delta+\epsilon, \label{outliermod}
\end{align}
where $y=(y_1,\dots,y_M)^T$ is an $M$-vector of responses, $\beta_0$ is an intercept term, $1_M$ denotes an $n$-vector of ones, 
$\beta$ is a $E\times 1$ vector of regression coefficients, $\delta=(\delta_1,\dots,\delta_M)^T$ is an $M$-vector of mean shift parameters
intended to be sparse and which allows for outliers in a small number of observations, and $\epsilon\sim N(0,\sigma^2 I)$.  
The model is not identifiable unless sparsity assumptions are made for $\delta$, and in the case where $E>M$,
which is the case we consider here, we also need to make some assumptions of sparsity for $\beta$.  

We consider a Bayesian analysis with priors $\beta_0\sim N(0,\sigma_0^2)$ and $\sigma\sim \mbox{HC}(0,A_\sigma)$ (where
$\mbox{HC}(0,A_\sigma)$ denotes the half Cauchy distribution with scale parameter $A_\sigma$).  The elements of $\beta$ are independent in their prior, 
$\beta_e\sim N(0,\sigma_e^2)$, with $\sigma_e\sim \mbox{HC}(0,A_\beta \gamma_e)$, $\gamma_e\sim \mbox{HC}(0,1)$, $e=1,\dots, E$, and $A_\beta$ is a scale parameter
to be chosen.  Similarly in the prior for $\delta$ the elements of $\delta$ are independent in the prior with 
$\delta_m\sim N(0,\tau_m^2)$, $\tau_m^2\sim \mbox{HC}(0,A_\delta \zeta_n)$, $\zeta_m\sim \mbox{HC}(0,1)$ for $m=1,\dots, M$, where
$A_\delta$ is a hyperparameter to be chosen.   
The prior specification is complete once the hyperparameters $\sigma_0^2$, $A_\sigma$, $A_\beta$ and $A_\delta$ are fixed.  In the current
section we consider the model where $\delta=0$ and hence there is no need to set $A_\delta$ and where $\sigma_0^2$ is fixed at $100$.  The full model is considered
further in the next subsection.  

We consider choice of $(A_\sigma,A_\beta)$ in the context of the sugar data set considered in \shortciteN{brown+vf98}.  In this dataset there are
$E=700$ predictors in the training sample, 3 response variables and 125 observations in the training set, so that we are considering a case where $E>M$.  We consider
the response variable glucose and center and scale all columns of the design matrix.  Now consider applying our method.  For summary statistics, we define
$S^1$ to be the log of the marginal variance of $y$ averaging over the predictors, i.e.\ $S^1=\log s^2$ where 
$$s^2=\frac{1}{n-1}\sum_{j=1}^n (y_j-\bar{y})^2,$$
where $\bar{y}$ is the sample mean of $y$.  We take the log in defining $S^1$ since $s^2$ can have quite a heavy tailed prior predictive distribution
due to the half-Cauchy prior on $\sigma$.  
Some idea of the range of the responses marginally is very likely to be available in applications and so it may be
easy to specify what would be plausible or implausible values for $S^1$.  We consider $S^1=\log 16$ to be plausible and 
$S^1=\log 50$ to be implausible (the marginal variance for the observed data is about 16 here).  

We also consider another summary statistic $S^2=S^2(y)$ defined as follows.  This summary statistic is an adjusted $R^2$ type measure of how much variation is
explained by the predictors, but one that is appropriate to the situation of more covariates than observations and which is based on a simple version of
the refitted cross-validation method of \shortciteN{fan+gh12}. 
Details of computation of this adjusted $R^2$ measure are given in the Appendix.  
We want to require that both $S^2=0.05$ 
as well as $S^2=0.95$ are plausible, so that
the model allows both a small or large amount of variation in the response variable to be explainable through the regression {\it a priori}.  

\begin{figure}
\begin{center}
\begin{tabular}{cc}
\includegraphics[width=70mm]{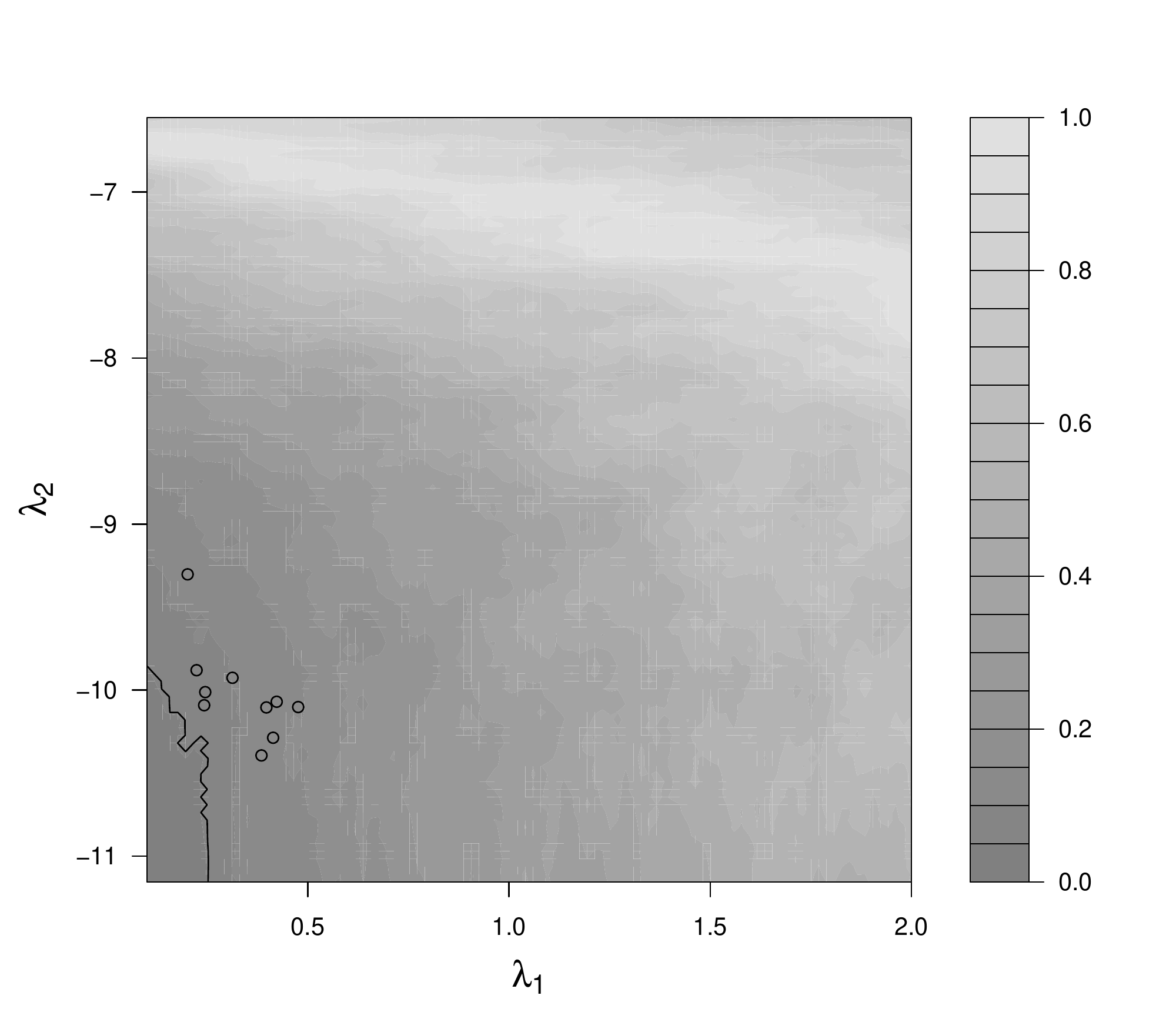}  & 
\includegraphics[width=70mm]{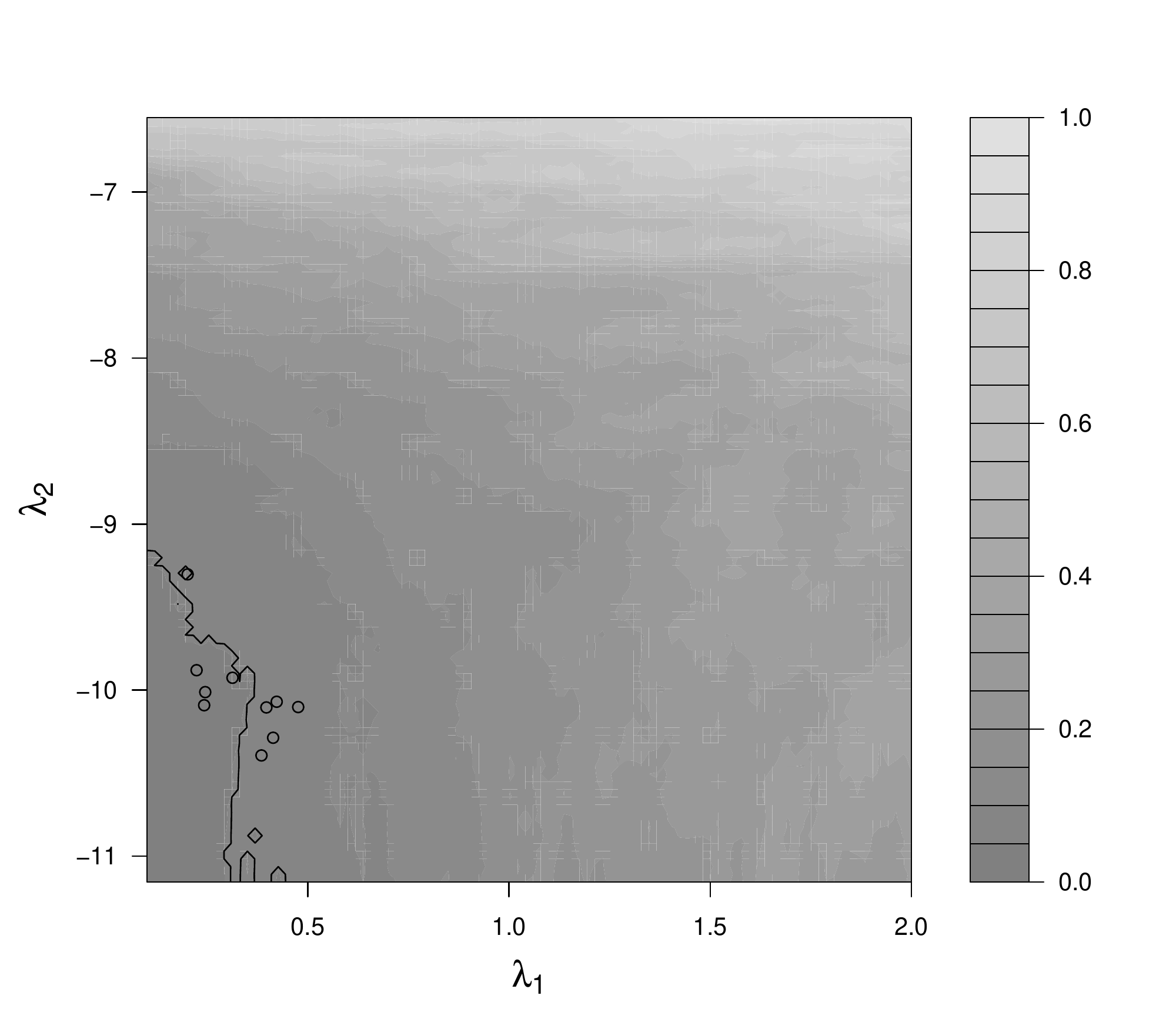} \\
\includegraphics[width=70mm]{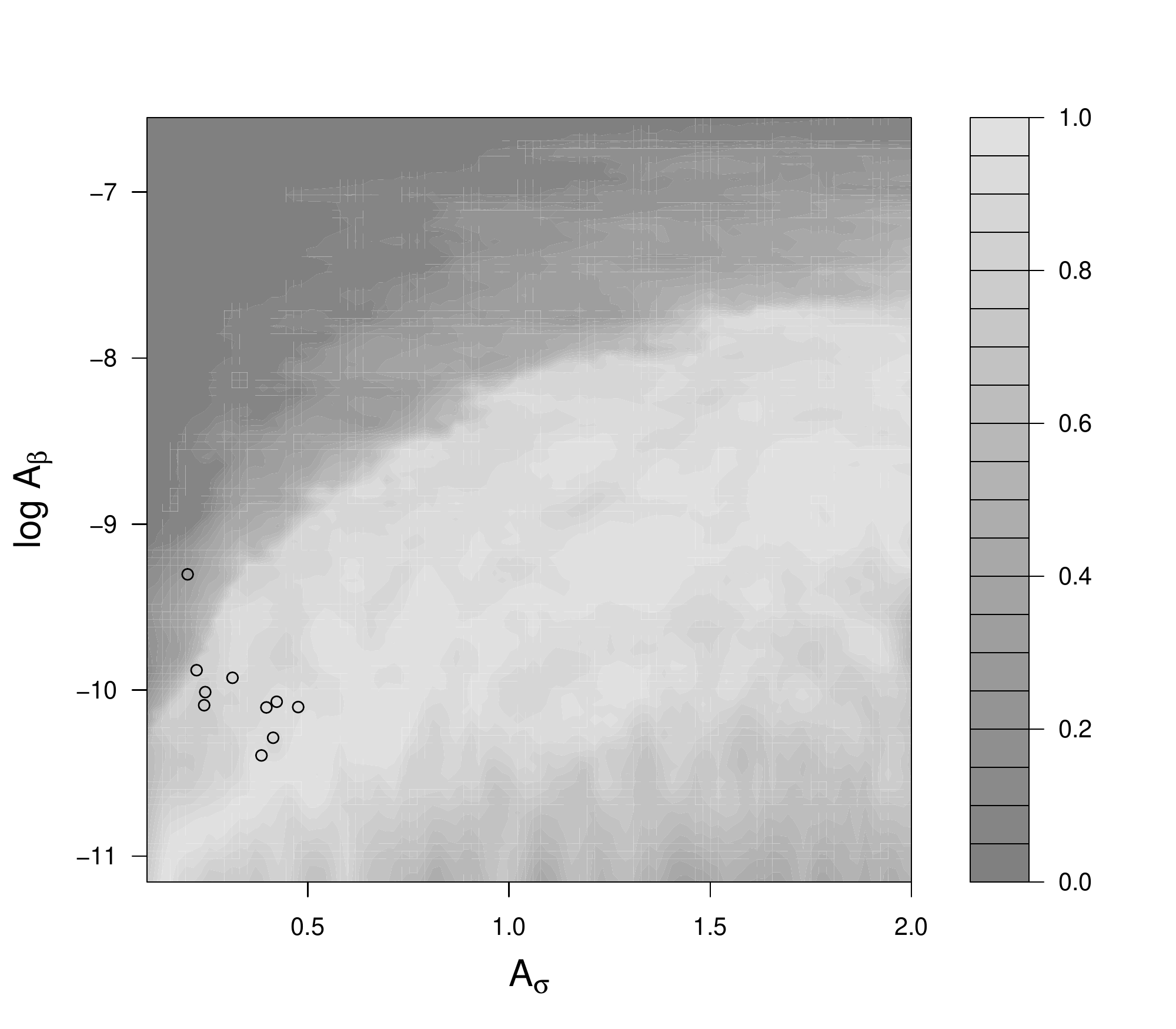}  & 
\includegraphics[width=70mm]{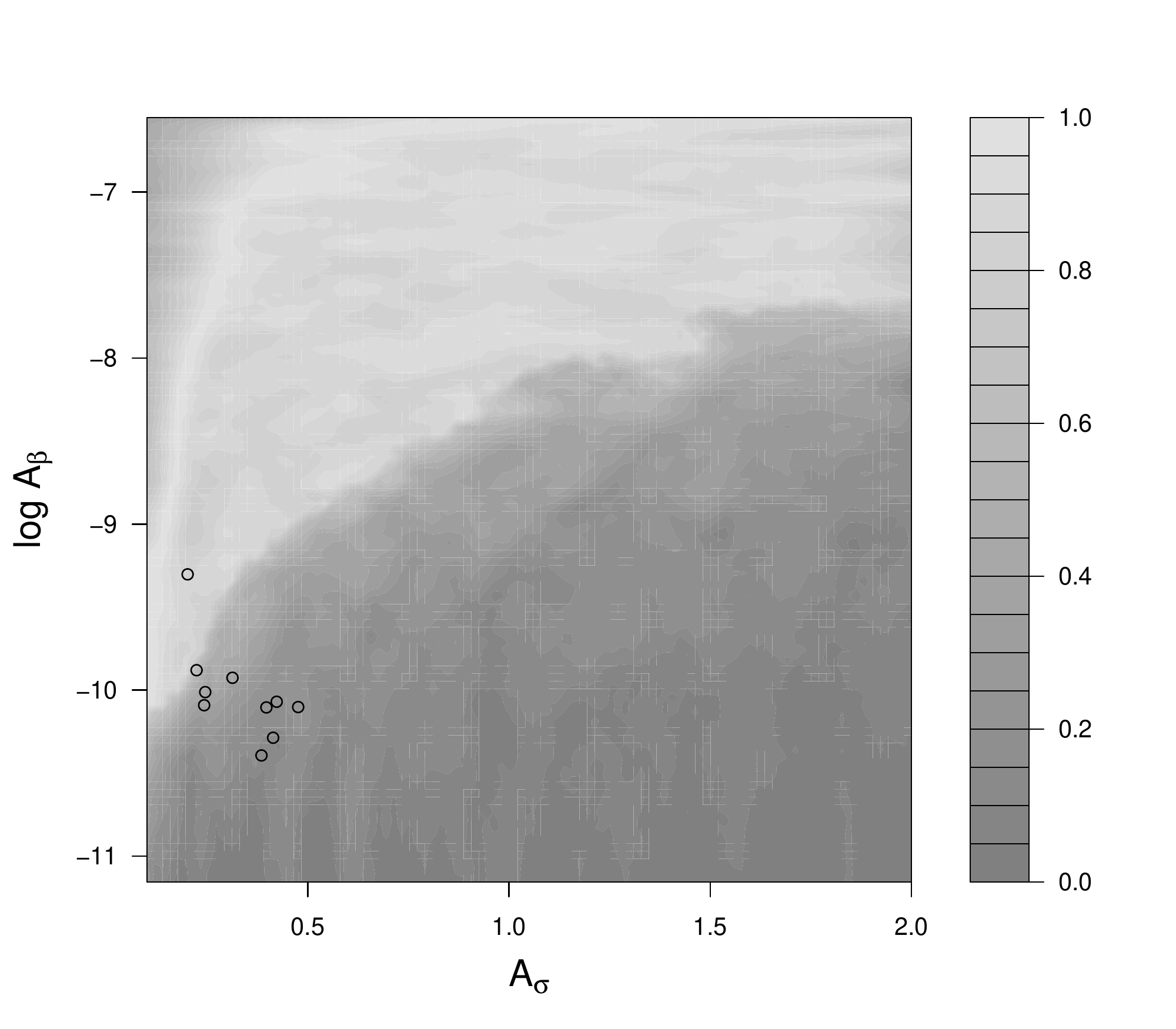} \\
\end{tabular}
\end{center}
\caption{\label{pvalueplot2} 
Conflict $p$-value as a function of $(A_\sigma,A_\beta)$ for sparse signal shrinkage example. $p$-value for check for $S^1=\log 16$ (top left), $S^1=\log 50$ (top right),
$S^2=0.05$ (bottom left) and $S^2=0.95$ (bottom right).  
In both graphs the overlaid points are from the third wave of the history match and the minimum implausibility obtained is zero. In the panels in the top row the contour line is at the level $0.05$.}
\end{figure}

Figure \ref{pvalueplot2} shows plots of the $p$-values for the tests based on the four summary statistics as $(A_\sigma,A_\beta)$ vary.   
The plots are for $100\times 100$ grids equally spaced in each dimension for $(A_\sigma,\log A_\beta)$ covering the range $[0,2]\times [-\log 100p,-\log p]$.  
The regression adjustment calculations for computation of the $p$-values are done using 100,000 evaluations of the summary statistics with local
linear regression adjustments and 1,000 nearest neighbours.
Similar to the last example overlaid on the graphs are the retained points from the third wave of a history match implemented in the same way as the previous
example with $r=100$ and $\gamma=0.1$.  The history match succeeds in finding prior hyperparameter values corresponding to priors which satisfy the constraints.
In the top right plot we want to be in the darkest grey region (i.e. the corresponding summary is implausible), 
and in the other plots we want to avoid the darkest grey region (i.e. the corresponding summaries are plausible).  In the top two panels in Figure 3 
the contour line is at the level $0.05$, showing
we have succeeded in finding points satisfying the constraint.

\begin{figure}
\begin{center}
\begin{tabular}{cc}
\includegraphics[width=70mm]{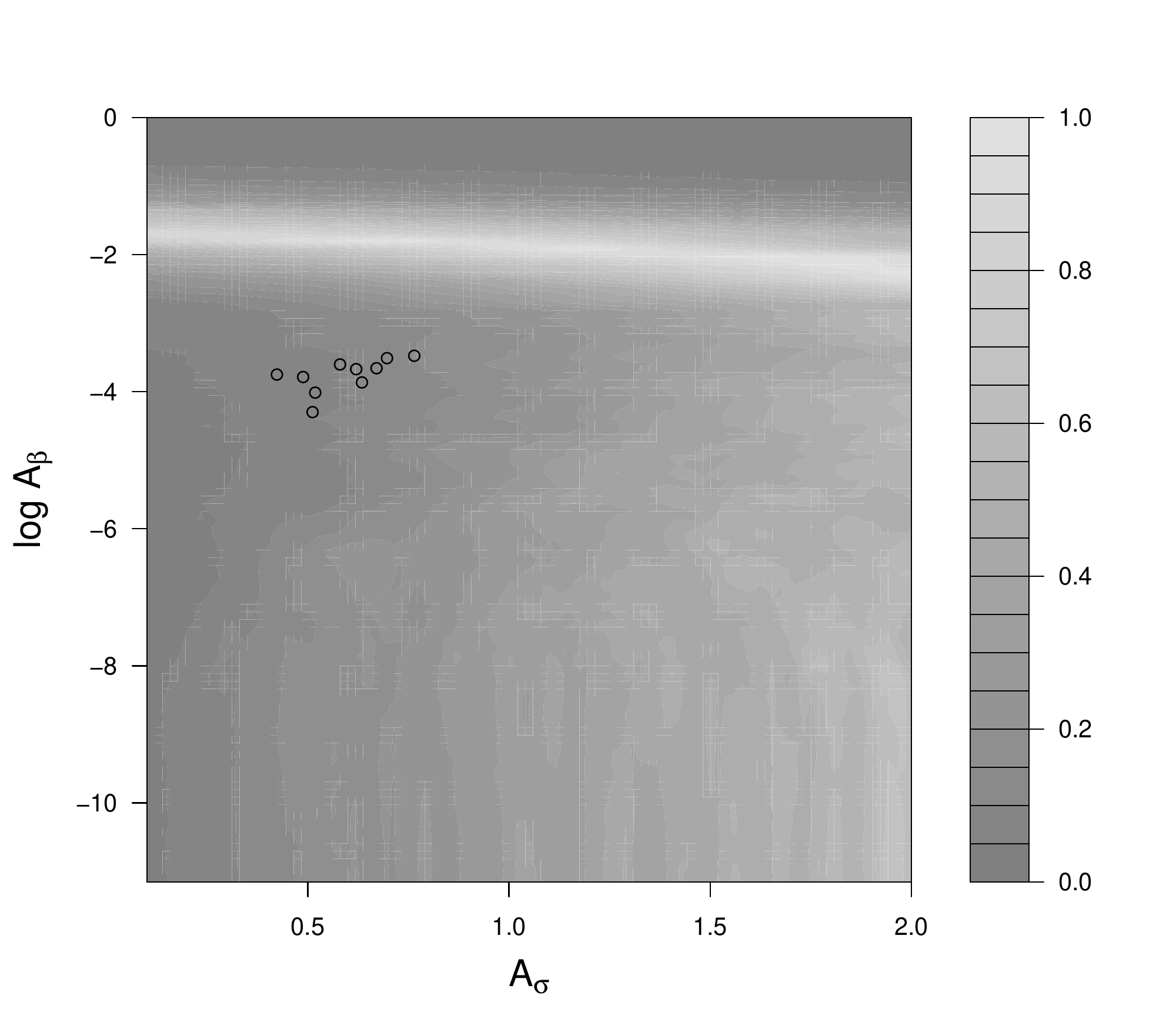}  & 
\includegraphics[width=70mm]{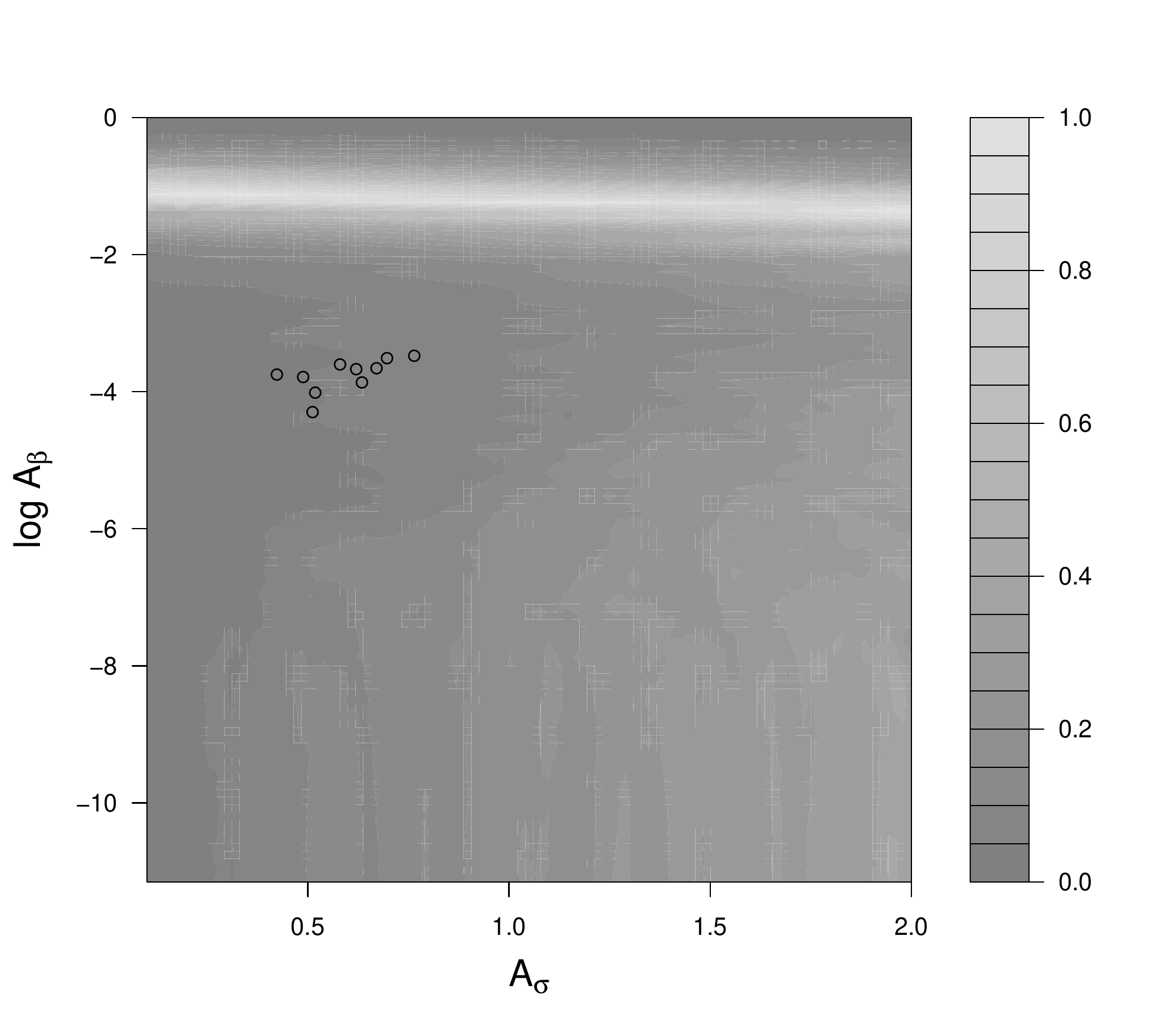} \\
\includegraphics[width=70mm]{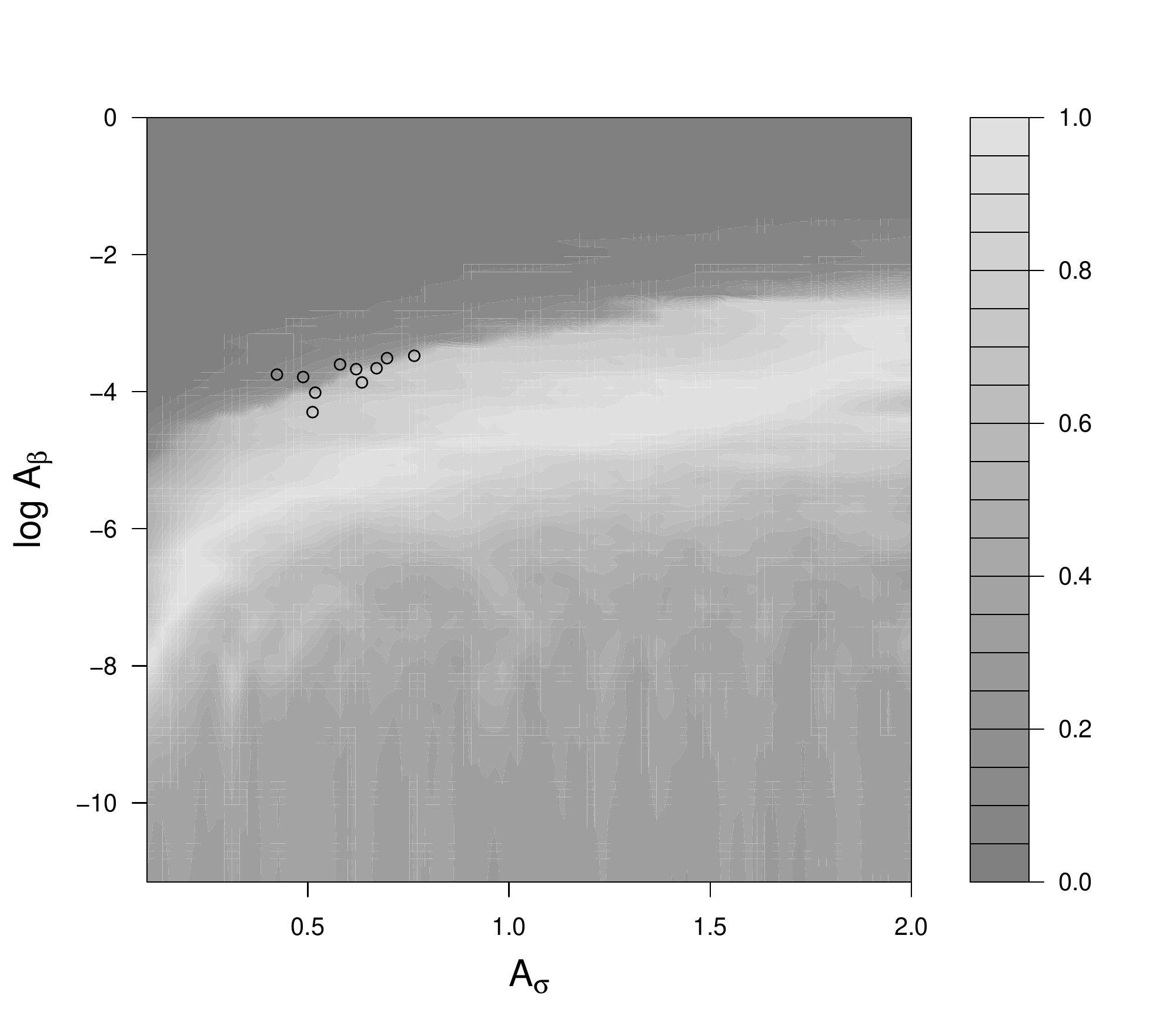}  & 
\includegraphics[width=70mm]{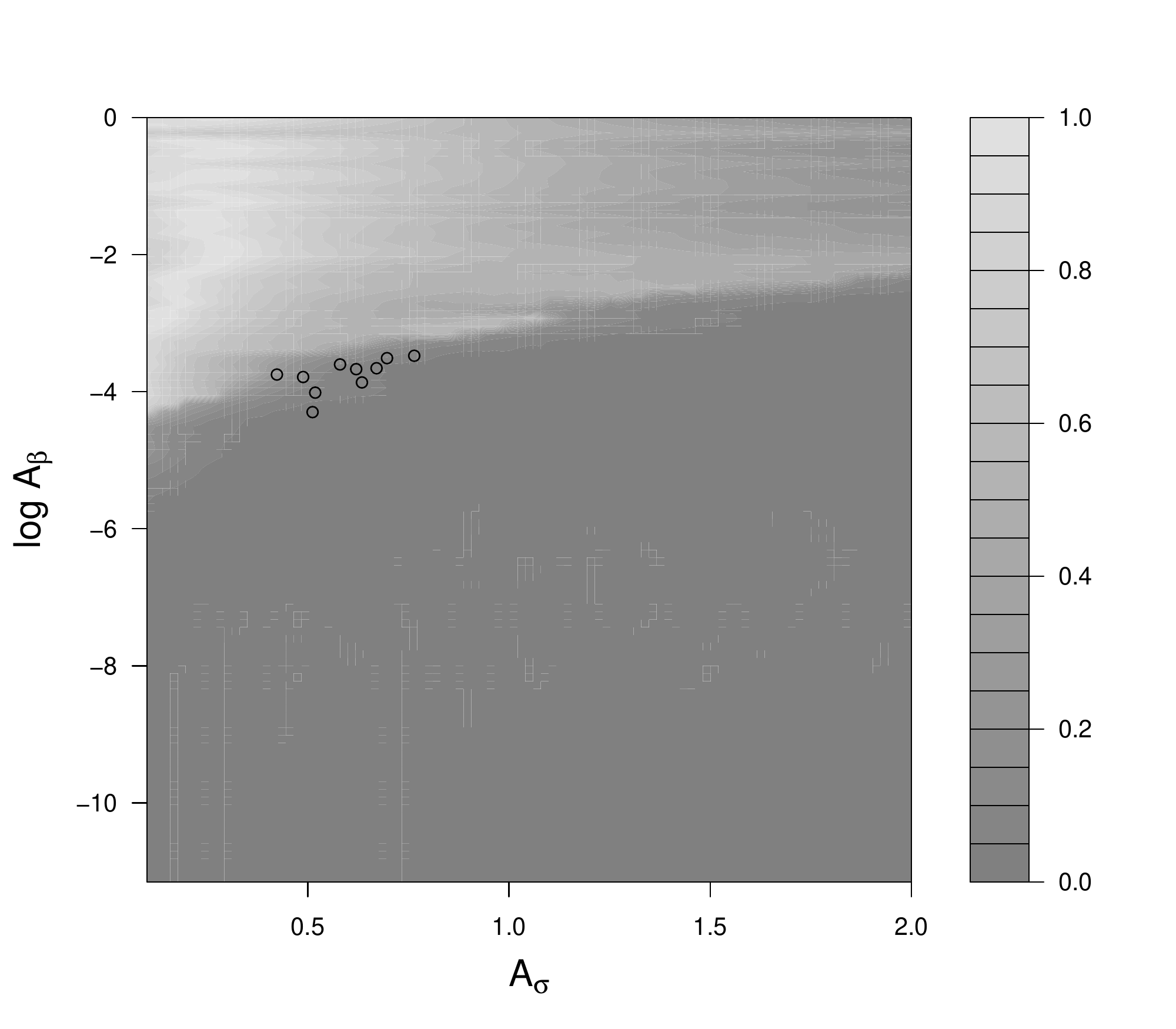} \\
\end{tabular}
\end{center}
\caption{\label{pvalueplot3} 
Conflict $p$-value as a function of $(A_\sigma,A_\beta)$ for normal prior example. $p$-value for check for $S^1=\log 16$ (top left), $S^1=\log 50$ (top right),
$S^2=0.05$ (bottom left) and $S^2=0.95$ (bottom right).  
In both graphs the overlaid points are from the third wave of the history match and the minimum implausibility obtained is 0.}
\end{figure}

It is interesting to see what happens in this example when we change the prior on $\beta$ to $\beta_j\sim N(0,A_\beta)$, so that now $A_\beta$ is a scale parameter
to be chosen in a normal prior, but where our predictive constraints remain the same.  We continue to use the notation $A_\beta$ for the scale parameter in the prior on $\beta$ 
even though this is of course a different parameter in the two priors.  
State of the art sparsity inducing priors like the horseshoe+ have good frequentist performance in a number of senses as described in \shortciteN{bhadra+others2015}.  
Here we illustrate a more Bayesian way in which this prior is good in this example.  
Before we did a history match in this example we
expected that the normal prior would work poorly in the sense of not being able to capture the information
that either a large or small amount of the variation in the response should be explainable through the covariates {\it a priori}.  Our intuition was incorrect, and
it was in fact possible to satisfy our constraints.  
The results of wave 5 of our history match for the normal prior are shown in Figure \ref{pvalueplot3}. 

\begin{figure}
\begin{center}
\begin{tabular}{cc}
\includegraphics[width=70mm]{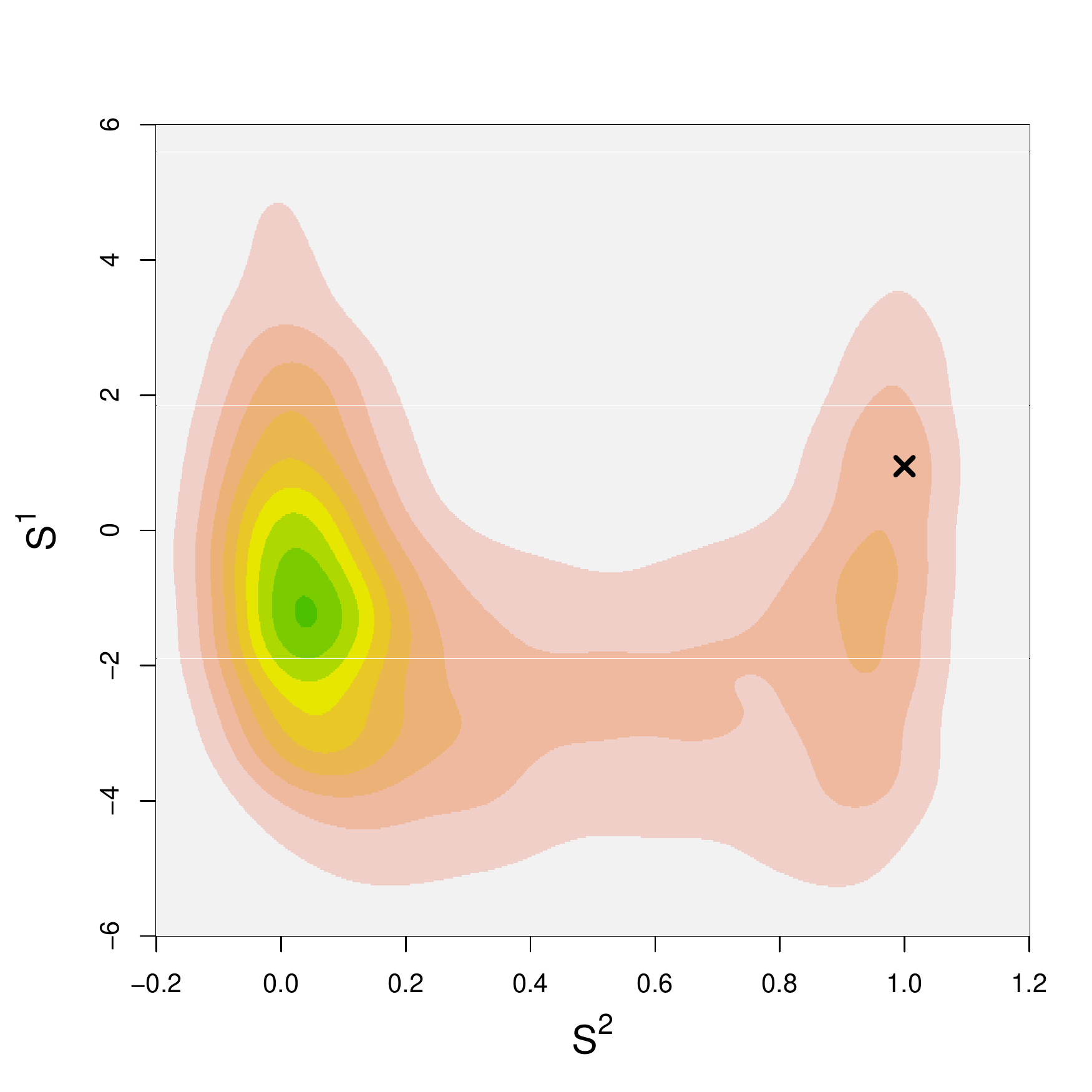}  & 
\includegraphics[width=70mm]{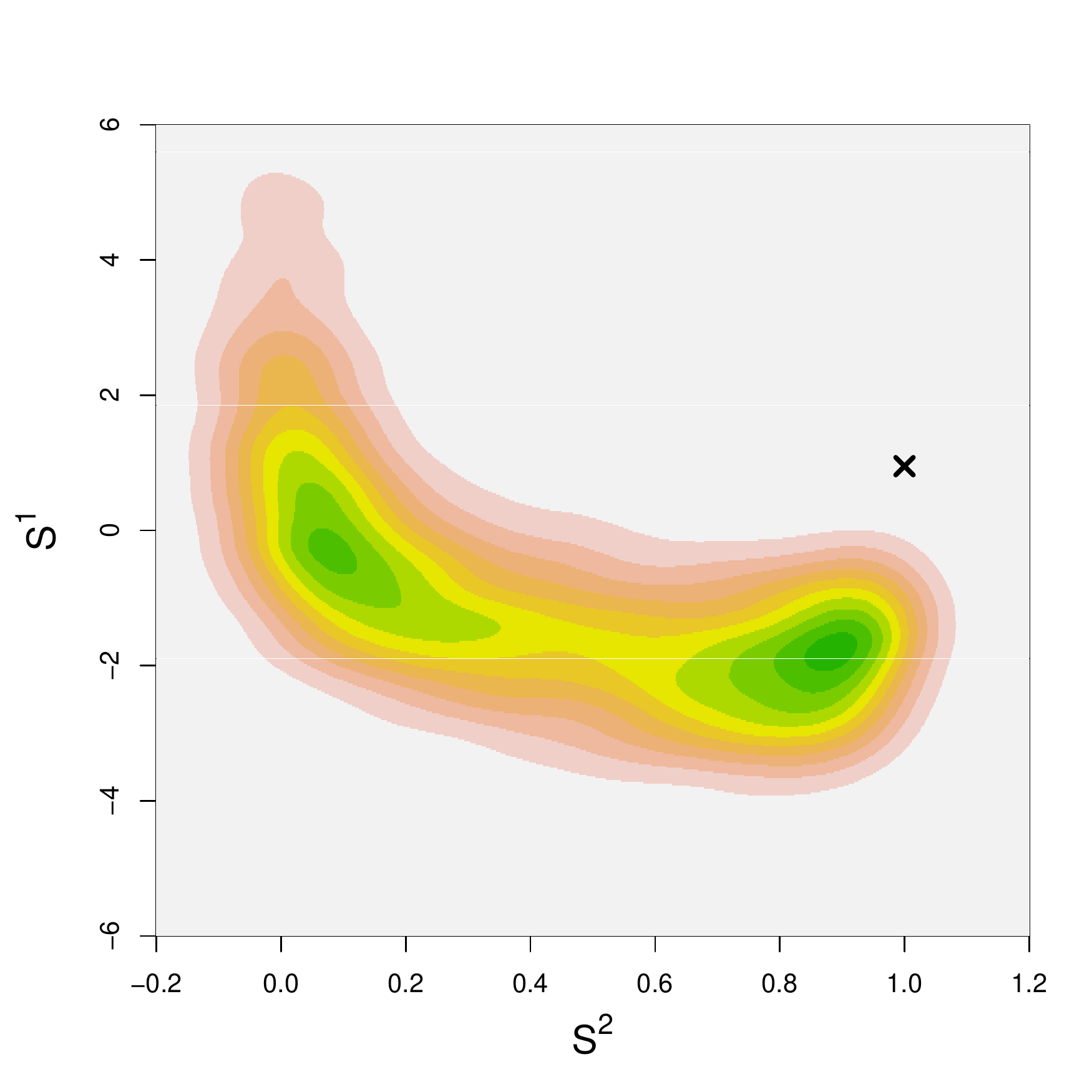} 
\end{tabular}
\end{center}
\caption{\label{densityplots2d} 
Prior predictive densities for $(S^1,S^2)$ for two zero implausibility hyperparameter values for horseshoe+ prior (left) and normal prior (right).  The point 
$(S^1,S^2)=(\log 16,0.95)$ is marked. The hyperparameters are $(A_\sigma, A_\beta) = (0.36, 0.014)$ for the normal prior, and 
$(A_\sigma, A_\beta) = (0.033,0.00004)$ for the horseshoe+ prior.  The same scale is used for both contour plots.  }
\end{figure}

However, now consider the following.  If $S^1=\log 16$ and $S^2=0.95$ should both be plausible, perhaps we should also require that $(S^1,S^2)=(\log 16,0.95)$ should be  plausible in the joint prior predictive for $(S^1,S^2)$.  Figure \ref{densityplots2d} shows kernel estimates of the joint prior predictive density
for $(S^1,S^2)$ for the horseshoe+ and normal priors for two particular hyperparameter values achieving zero implausibility, based on 1000 prior predictive samples.  
We can see that $(S^1,S^2)=(\log 16,0.95)$ is plausible for the horseshoe+ prior, but not for the normal prior.  
The explanation for this is that it is only when the noise variance is small that the regression can explain a lot of the variation in the case of the normal prior.  The behaviour of 
the horseshoe+ prior, however, is more acceptable.  
This example illustrates perhaps some of the pitfalls of considering plausible and implausible values for one-dimensional summary statistics separately.  While this is a useful
strategy for defining constraints, and it makes computations more convenient, once a reasonable candidate hyperparameter value is found it may be useful to consider the behaviour
of the joint prior predictive for several summaries simultaneously.

\subsection{An example with higher-dimensional hyperparameter}

\begin{figure}
\begin{center}
\begin{tabular}{c}
\includegraphics[width=150mm]{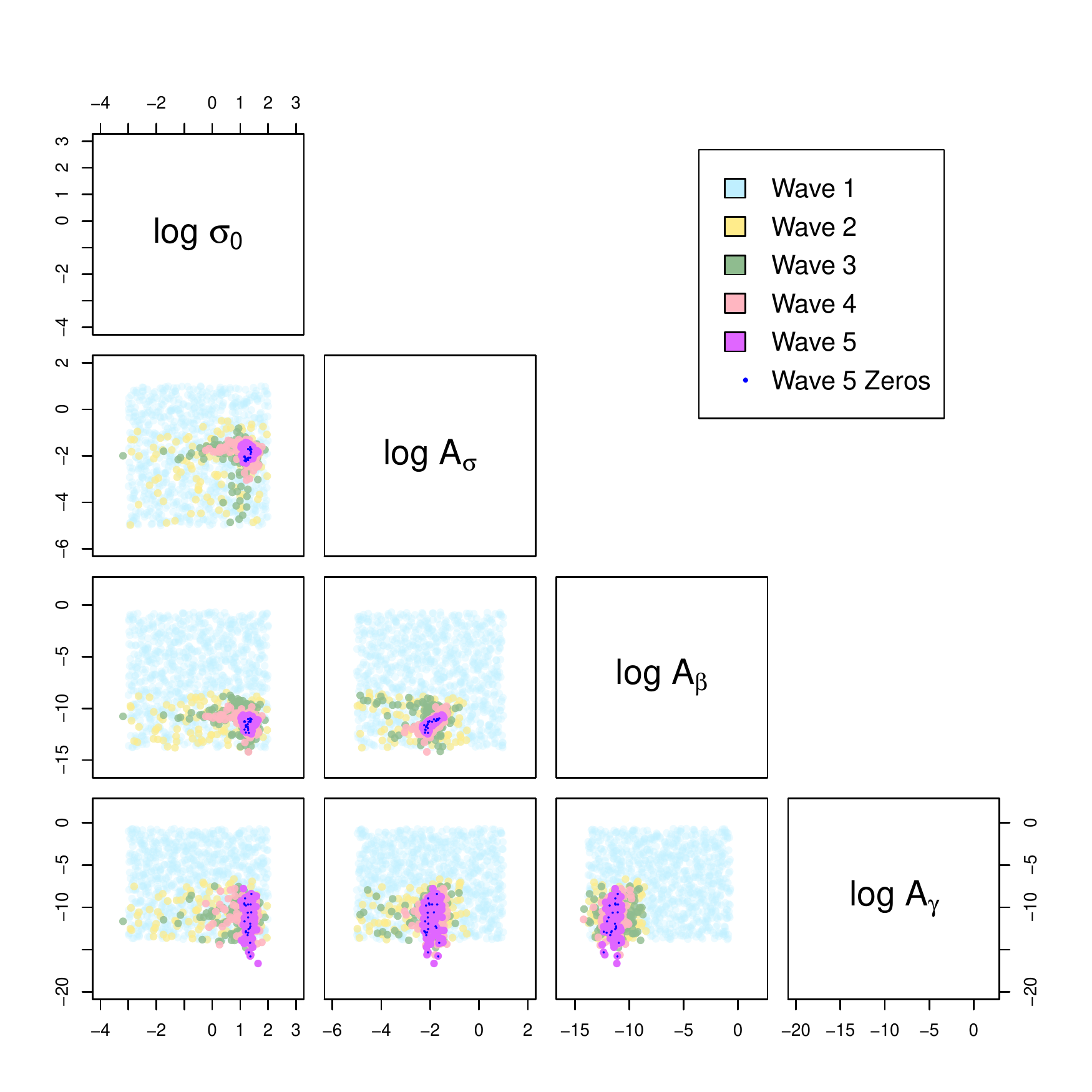} 
\end{tabular}
\end{center}
\caption{\label{hmatchplot} 
Pairwise scatterplots of hyperparameters on log scale of wave 1 to wave 5 of the history match. The minimum implausibility value obtained in wave 5 is 0.}
\end{figure}

Continuing the last example, consider the full model (\ref{outliermod}) described in Section 5.2 where now we allow $\delta$ to be nonzero.  We also consider
the situation where $\sigma_0^2$ is not fixed in the prior for $\beta_0$.  Now we have four hyperparameters to be chosen, $(\sigma_0,A_\sigma,A_\beta,A_\delta)$.
Unlike the previous two examples with only two hyperparameters, it is not feasible to use a grid-based approach to produce plots of how the conflict $p$-values
vary over the hyperparameters for comparison with the results of the history match.  We retain the summary statistics and constraints of Section 5.2, with the difference
that $s^2$ is replaced by a robust measure of scale (the median absolute deviation estimator), and in the linear regression fits for the refitted cross-validation procedure we use the robust 
{\tt lmrob} function in {\tt R} (\shortciteNP{rosseuw+ctrsvkm15}) to obtain
the adjusted $R^2$ estimate. We also add to the constraints of Section 5.2 three additional constraints.  
We choose a summary statistic $S^3$ to be the log of the absolute value of the median of the responses, 
and specify $S^3=\log 15$ to be plausible, and $S^3=\log 20$ to be implausible.  
As an additional summary statistic we use the following procedure.  We consider the log sample kurtosis of the residuals obtained from the {\tt lmrob} function
averaged over 10 split samples using the same refitted cross-validation procedure as for the adjusted $R^2$ measure.  This is intended to be some sample measure
of the ``tailedness" of the distribution.  
Writing $S^4$ for this statistic, we consider $S^4=\log 50$ to be implausible.  The value of $\log 50$ was obtained as the log of the approximate median
of sample kurtosis values from a Cauchy distribution sample of size $125$.  Note that we use sample kurtosis here as a summary of the data without worrying
about whether any corresponding population quantity exists.  The information in this last summary statistic is intended to state the requirement that we should not have
a very large proportion of very extreme outliers.
Figure \ref{hmatchplot} shows pairwise scatter plots of the hyperparameter values on a log scale in wave 1 through wave 5 of a history match 
with $r=1000$ and $\gamma=0.1$ and the first wave initialized with a maximin latin hypercube
design covering the range $[e^{-3},e^{2}]\times [e^{-5},e] \times [10^{-6},0.5] \times [10^{-6},0.5]$ for the hyperparameters.  
The history match succeeds in finding prior hyperparameter values corresponding to priors which satisfy the constraints.

\begin{figure}
\begin{center}
\begin{tabular}{c}
\includegraphics[width=150mm]{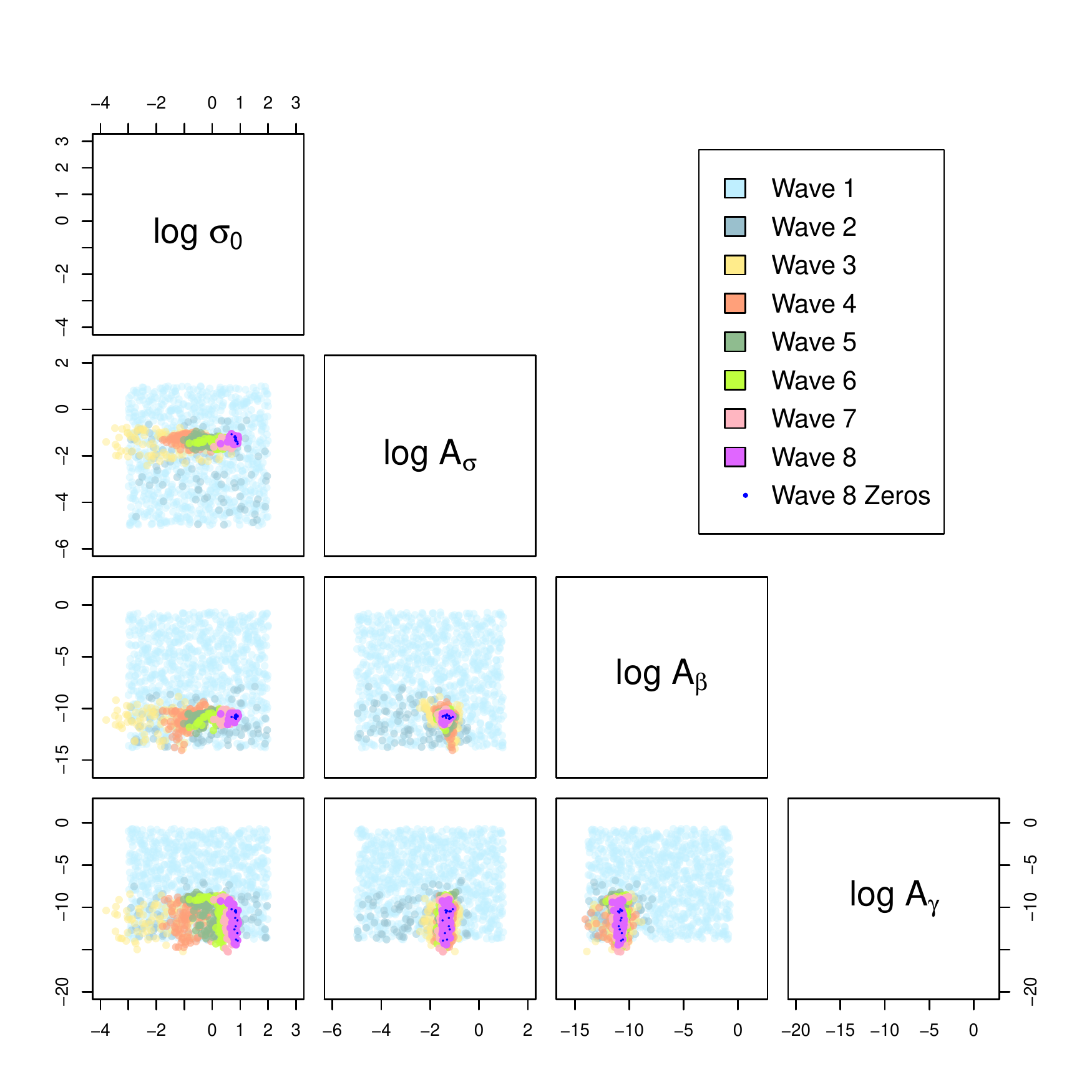} 
\end{tabular}
\end{center}
\caption{\label{hmatchplot2} 
Pairwise scatterplots of hyperparameters on log scale of wave 1 to wave 8 of the history match with additional model simulations at each wave. The minimum implausibility value obtained in wave 8 is 0.}
\end{figure}

\begin{figure}
\begin{center}
\begin{tabular}{cc}
\includegraphics[width=70mm]{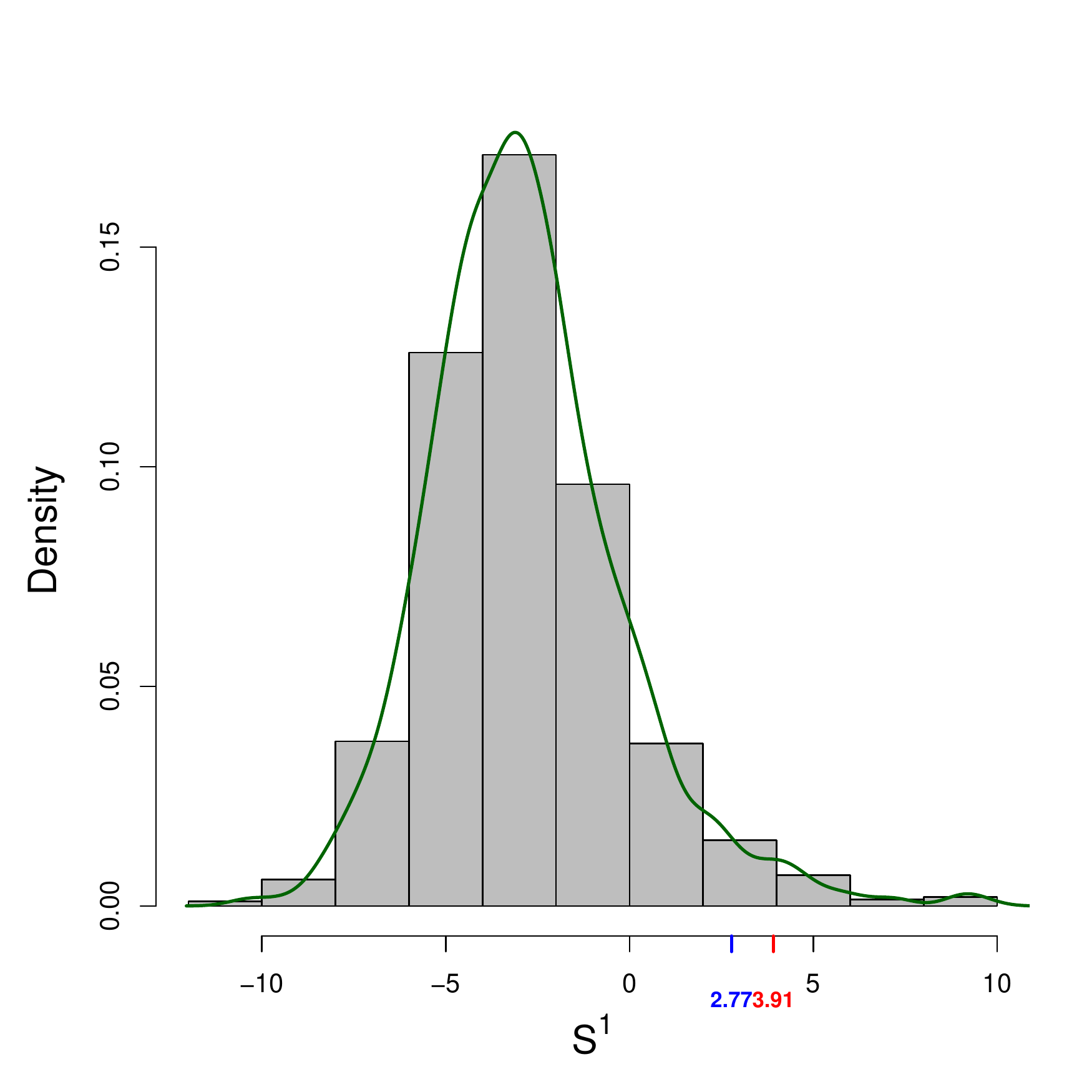}  & 
\includegraphics[width=70mm]{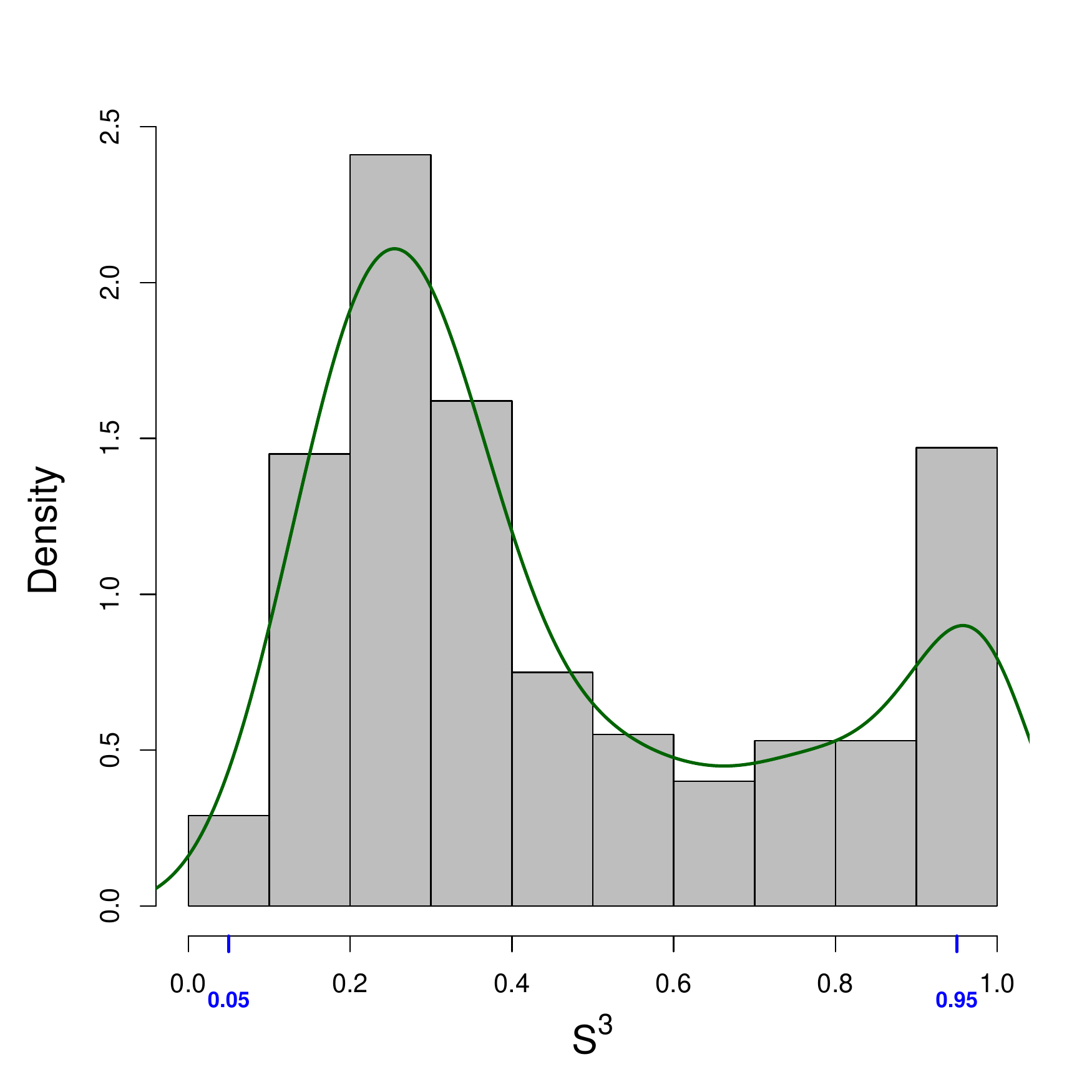} \\
\includegraphics[width=70mm]{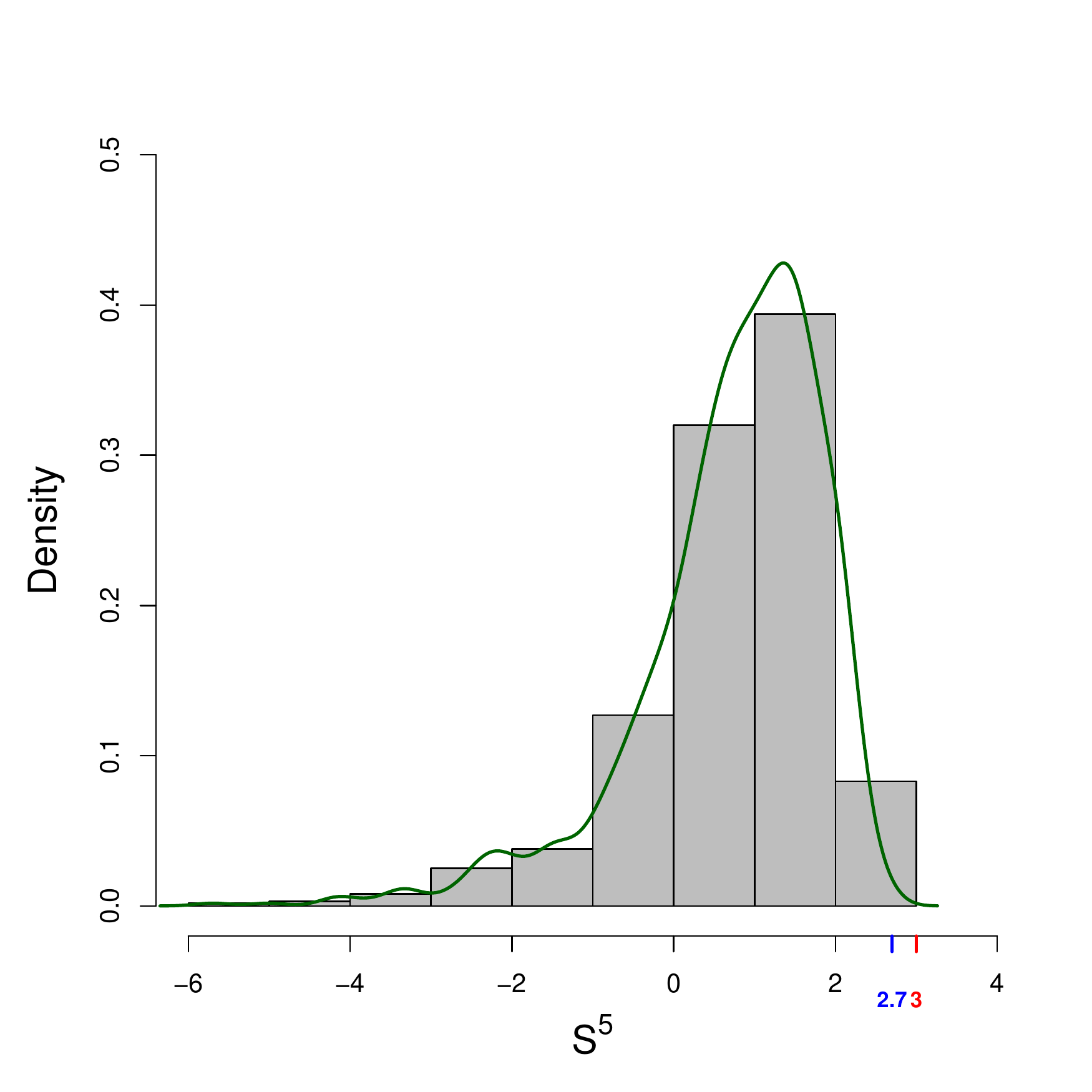}  & 
\includegraphics[width=70mm]{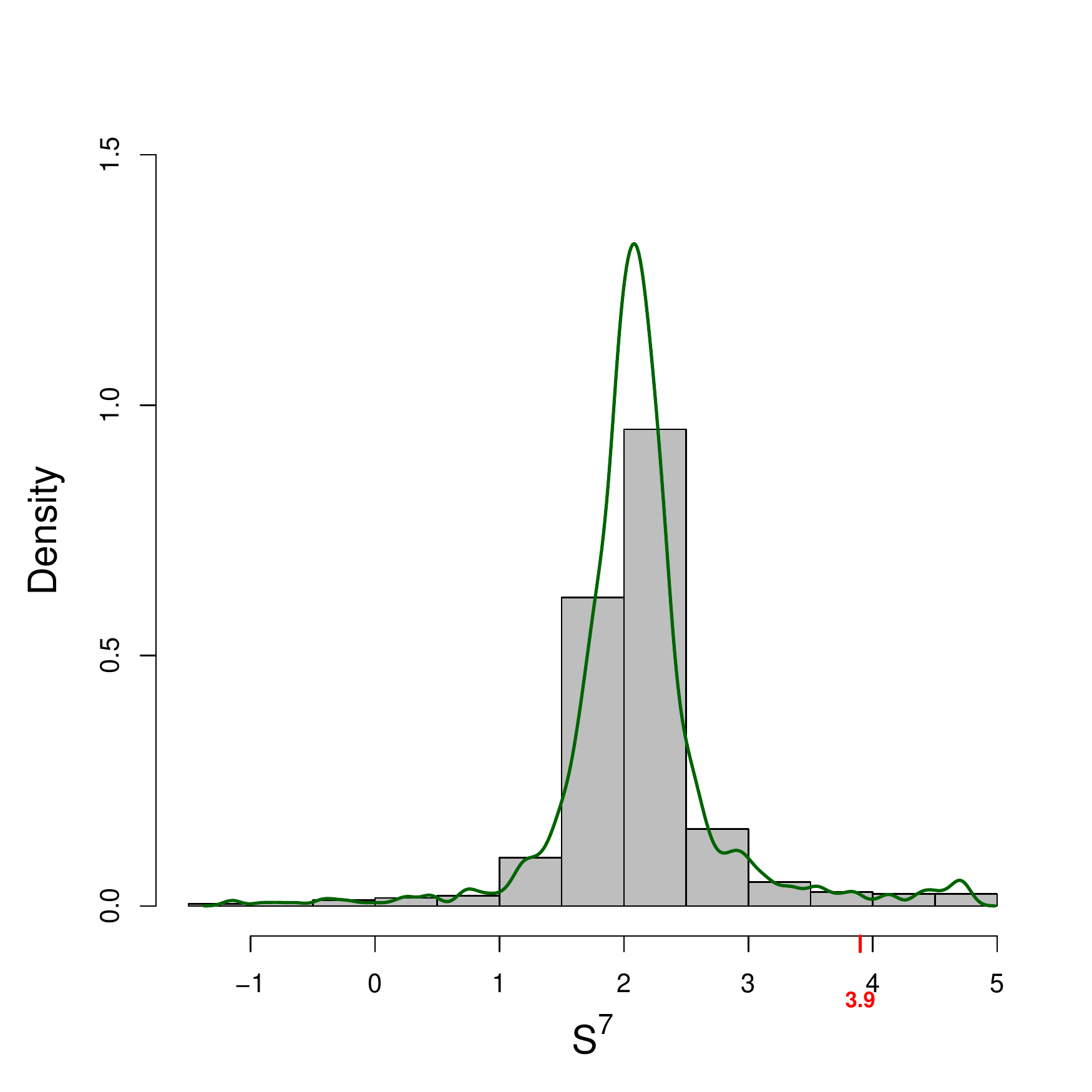} \\
\end{tabular}
\end{center}
\caption{\label{predictive} 
Prior predictive densities of $S^1$, $S^2$, $S^3$, $S^4$ for hyperparameter value achieving zero implausibility.  The red and blue numbers are the plausible (blue) and implausible (red) hypothetical values for the summaries used in the history match.}
\end{figure}

In Section 4 it was mentioned that it may be helpful to adaptively generate new summary statistic simulations as the waves of the history match proceed.  The results
of Figure \ref{hmatchplot} were obtained without doing this, using $100,000$ simulations at the beginning of the procedure.   Figure \ref{hmatchplot2} 
shows 8 waves of a history match where
the initial number of summary statistic simulations was reduced to $10,000$, with $1,000$ additional simulations added at each wave ($100$ further simulations at each of the $10$ non-implausible values retained at each wave).  The results are similar to before, but now the total number of model simulations has been reduced to $18,000$ rather than $100,000$.  
Although this is not a very high-dimensional example, this illustrates the point that this adaptive approach to the model simulations to improve the quality of the regression ABC
adjustment can be very important as the number of hyperparameters increases.  Effectively the additional model simulations allow us to use smaller neighbourhoods in this local
nonparametric procedure.  Any approach to flexible conditional density estimation could be used instead of the regression ABC approach for approximating the prior predictive
densities as a function of the hyperparameters, but any such alternative method will also
benefit from additional model simulations in the important region of the space.  
Figure \ref{predictive} shows estimated prior predictive densities of the summary statistics used in the history match obtained from one of the hyperparameter values 
with implausibility zero in Figure \ref{hmatchplot}, 
$(\sigma_0, A_\sigma, A_\beta, A_\gamma) = (3.91, 0.016,0.000013, 0.000045)$.   
The graphs presented are histograms and kernel density estimates based on 1000 prior predictive samples. 

\section{Discussion}

We have considered a novel application of the ideas of history matching used in the assessment of computer models to the problem
of prior choice.  By defining the implausibility measure in the history match through some prior predictive constraints, we are able to implement predictive
elicitation even for complex models.  
Regression adjustment ABC methods are also used to ease the computational burden in application of the method.  
We believe the analyses presented in some of the examples are insightful, and in some cases led to some new understanding of the effects of the parameter prior
on the prior predictive densities.

Further investigation is needed to see how well the methods we have developed scale to problems where the number of hyperparameters is 
much larger.  Also, it is not clear whether the specific form for the implausibility measure that was chosen was the best one.
Although, as we have stressed throughout the manuscript, we are focusing mostly on computational questions in this paper 
it is also worth considering how the 
methods and algorithms developed are best integrated within an elicitation procedure in complex applied problems.  

As noted in the introduction, while in this work we specify constraints in the form of passing or failing model checks for hypothetical
data, the constraints could also be specified in some other way in our procedure, such as through inequalities on quantiles of predictive distributions.  The numerical
search procedures developed later can also be used with constraints in these other forms.    
Our method can also apply in situations where prior information is expressed directly on the parameter itself rather than predictively.  
It is not uncommon for prior distributions to be specified conditionally through a hierarchy, and for marginal prior distributions for functions of the parameter
to be unavailable analytically.  We can consider tail probabilities for such marginal priors or inequalities on quantiles for such priors in the same
basic framework as our predictive methods.  Again, indicator functions for certain sets such as expressing order constraints on certain parameters might
be one useful way of adding information.  
The ABC computations in our method are similar to those used in \shortciteN{nott+dme15} for finding weakly informative priors and 
many of the elicitation calculations can be reused for finding such a weakly informative prior in the event that there is a prior-data conflict.  
Also worthy of further investigation is whether greater use can be made of the full set of prior distributions returned by
the history match.  Here we have simply focused on choice of a single ``adequate" prior but there is 
a richer source of information that can be used in the results of the history matching procedure.

\subsection*{Acknowledgements}
%%%%%%%%%%%%%%%%%%%%%%%%%%%%%%%
%%%%%%%%%%%%%%%%%%%%%%%%%%%%%%

David Nott was supported by a Singapore Ministry of Education Academic Research
Fund Tier 2 grant (R-155-000-143-112).  Christopher C Drovandi was supported by an Australian Research Council’s Discovery Early Career 
Researcher Award funding scheme (DE160100741).  Kerrie Mengersen was supported by an Australian Research Council Laureate Fellowship.
Michael Evans was supported by a grant from the
Natural Sciences and Engineering Research Council of Canada.

\bibliographystyle{chicago}
\bibliography{historymatchingpriorchoice}

\subsection*{Appendix}

We outline here the calculation of the adjusted $R^2$ type measure used as a summary in the example of Section 5.2.  A refitted cross-validation approach (\shortciteNP{fan+gh12})
is used based on $10$ random splits.  
The algorithm is as follows.
\begin{enumerate}
\item For $j=1,\dots, 10$, 
\begin{enumerate}
\item Split the data $y$ into two halves, $y=({y^{(1)}}^T,{y^{(2)}}^T)$.  Split $X$ similarly as $X=[{X^{(1)}}^T\; {X^{(2)}}^T]^T$.  
\item Compute the absolute value of the Pearson correlation of $y^{(1)}$ with column $i$ of $X^{(1)}$.  Write this as $R^{i,j,1}$, $i=1,\dots,  p$.  
Similarly compute the absolute value of the Pearson correlation of $y^{(2)}$ with column $i$ of $X^{(2)}$ and write this as $R^{i,j,2}$, $i=1,\dots, p$.  
\item Let $S^*(k)$ denote the indices $i$ of the predictors with the $M/4$ largest values of $R^{i,j,k}$, $k=1,2$.  
\item Write $X_{S^*}^{(1)}$ for the submatrix of $X^{(1)}$ which retains only columns $i\in S^*(2)$, and similarly $X_{S^*}^{(2)}$ is the submatrix
of $X^{(2)}$ which retains only columns $i\in S^*(1)$.  Fit a linear regression model
of $y^{(1)}$ on $X_{S^*}^{(1)}$ and write the adjusted $R^2$ for this regression as $R^{(j,1)}$.  Similarly fit a linear regression model of $y^{(2)}$ on $X_{S^*}^{(2)}$ 
and write the adjusted $R^2$ for this regression as $R^{(j,2)}$.  Write $R^{(j)}=0.5\times (R^{(j,1)}+R^{(j,2)})$.    
\end{enumerate}
\item $S^2(y)=\frac{1}{10}\sum_{j=1}^{10} R^{(j)}$
\end{enumerate}

\end{document}